\documentclass[conference]{IEEEtran}
\IEEEoverridecommandlockouts

\usepackage{cite}
\usepackage{amsmath,amssymb,amsfonts}
\usepackage{algorithm}
\usepackage{algorithmic}
\usepackage{graphicx}
\usepackage{textcomp}
\usepackage{xcolor}
\usepackage{hyperref}
\usepackage{array}

\DeclareMathOperator{\sign}{sign}
\newcommand{\RE}{\mathbb{R}}

\newtheorem{remark}{Remark}
\newtheorem{assumption}{Assumption}

\DeclareMathOperator{\Deriv}{d}
\DeclareMathOperator{\Real}{Re}
\DeclareMathOperator{\Imag}{Im}

\definecolor{GreenMatLab}{rgb}{0, 0.5, 0}

\begin{document}

\title{The Bode Plots for Sliding-Mode Control Design
}

\author{\IEEEauthorblockN{Ulises Pérez-Ventura}
\IEEEauthorblockA{\textit{Department of Control and Robotics, Division of
Electrical Engineering,} \\
\textit{National Autonomous University of Mexico,}\\
Mexico City, 04510,  \\
ventury.sk8@gmail.com}
\and
}

\maketitle

\begin{abstract}
This paper develops a unified frequency-domain framework for the analysis of sliding-mode control systems, encompassing both discontinuous and Lipschitz-continuous implementations. Using describing function (DF) theory, closed-form expressions are derived for the amplitude and frequency of chattering oscillations, as well as equivalent gain (EG) models that enable closed-loop sensitivity analysis. The proposed methodology captures the influence of actuator dynamics, control parameters, and disturbance profiles on steady-state performance.

Theoretical predictions for bias and oscillatory components are validated through simulations under both constant and sinusoidal perturbations. In the low-frequency regime, the EG-based sensitivity functions accurately predict the amplitude and phase of the system response, with tracking errors remaining within a 15\% margin, provided that the DF assumptions hold. The framework also incorporates orbital stability considerations via Loeb’s criterion, ensuring that chattering remains bounded.

Overall, the results offer practical insight into the robust design of sliding-mode controllers, enabling systematic gain tuning that balances disturbance rejection and chattering attenuation, while accounting for actuator and sensor constraints.
\end{abstract}


\section{Introduction}
The advent of low-cost computing platforms has solidified control systems as a foundational technology across a wide range of industrial domains. Automation now extends to fields such as healthcare, transportation, energy systems, and manufacturing. More generally, any dynamical process that can be described by a mathematical model may be subjected to analysis and synthesis via the tools of control theory \cite{Ogata02}. However, mathematical models of physical systems are frequently subject to parametric uncertainties, unmodeled dynamics, and external disturbances. Designing controllers that guarantee specified closed-loop performance in the presence of such non-idealities remains a central challenge in modern control theory. Sliding-Mode (SM) controllers have emerged as an effective methodology for the stabilization of uncertain dynamical systems subject to perturbations \cite{Utkin92}. Conceptually, SM algorithm aim to drive a suitably defined scalar function—commonly referred to as the sliding or switching variable—to zero in finite time  \cite{Shtessel14}. This variable may represent a tracking error, a state estimation error, a parameter identification error, or even the error arising from numerical differentiation of a signal with respect to time.

Robust solutions based on SM techniques have been extensively investigated in the literature for a variety of tasks, including automatic control  \cite{Emelianov86,Bartolini98,Feng02,Levant05a}, state observation \cite{Barbot09,Rios21}, and time differentiation \cite{Levant98,Angulo12}. In this context, High-Order Sliding Mode (HOSM) algorithms have attracted considerable attention due to their ability to preserve the robustness properties of classical SM approaches while mitigating undesirable chattering effects \cite{Chalanga13,Fridman15,Ding16,Laghrouche17,Cruz17,Moreno20}. The main features of HOSM-based methods include:
\begin{itemize}
\item[(i)] the ability to stabilize uncertain dynamical systems of arbitrary relative degree, ensuring finite-time convergence of the sliding variable to zero;
\item[(ii)] exact rejection of matched disturbances; and
\item[(iii)] high accuracy in the presence of unmodeled dynamics and measurement noise.
\end{itemize}
Although the \textit{theoretical advantages} of HOSMs are substantial, their practical implementation remains limited due to the lack of closed-loop performance guarantees \cite{Panathula17}. A major obstacle is the chattering phenomenon, manifested as high-frequency oscillations in the closed-loop signals \cite{Levant10}.  Chattering can significantly degrade the system performance and reliability, leading to mechanical wear, thermal stress in power electronics, and, in extreme scenarios, structural damage induced by resonance. Several techniques have been proposed to mitigate chattering effects, such as replacing the discontinuous control functions with smooth approximations—e.g., arctangent, saturation, or sigmoid-type functions \cite{Shtessel96,Castillo20}; using low-pass filters to suppress high-frequency components \cite{Lee07,Rosales15}; and implementing continuous HOSM algorithms \cite{Chalanga13,Fridman15,Ding16,Laghrouche17,Cruz17,Moreno20}. However, chattering fundamentally arises from the use of high (or even \textit{infinite}\footnote{The notion of infinite control gain refers to the fact that the slope of the control signal tends to infinity as the sliding variable approaches zero, implying an idealized discontinuity in the control action—at least locally.}) control gain in systems affected by parasitic dynamics \cite{Levant10}.

Two principal approaches have been developed to analyze and characterize chattering phenomena:
\begin{itemize}
	\item \textbf{Time-domain analysis}: An extension of \textit{singular perturbation theory} has been formulated for relay-controlled systems, providing sufficient conditions for the existence and stability of limit cycles in first-order systems \cite{Fridman01}. This framework was later generalized to second-order systems through the application of \textit{Poincaré maps} \cite{Boiko07b}. Nevertheless, the approach lacks scalability, as it cannot be readily extended to systems of order higher than two.

    \item \textbf{Frequency-domain analysis}: The \textit{locus of perturbed relay systems} technique enables exact computation of oscillation parameters in relay servo systems \cite{Boiko05,Boiko09}. In contrast, the Harmonic Balance (HB) method has proven effective in predicting the amplitude and frequency of main chattering harmonics, particularly in systems that satisfy the \textit{low-pass filter hypothesis}\footnote{This hypothesis assumes that the high-order harmonics generated by the nonlinear elements are effectively attenuated by the linear components within the closed-loop system \cite{Gelb68}.}. The HB approach has been widely used to analyze chattering in SM and HOSM control systems \cite{Boiko04,ByF05,Rosales17,Ventura18,Ventura19,Ventura21}, although its accuracy critically depends on the availability of an appropriate Describing Function (DF) that approximates the input–output behavior of the nonlinear control under sinusoidal excitation \cite{Gelb68}.
\end{itemize}






\subsection{Ideal Sliding-Mode}
Let $\sigma \in \mathbb{R}$ denote the \textit{sliding variable}, which typically represents an error signal—such as a tracking error, state estimation error, or parameter identification error—that the controller aims to nullify in finite time. Suppose that the error dynamics are dominated by a first-order behavior, which can be modeled as
\begin{equation}\label{CL}
\dot{\sigma}(t)=f(t)-u(t) \,, \vspace{-0mm}
\end{equation}
where $u(t)$ is the control input and $f(t)$ denotes an external disturbance. Consider the classical sliding-mode (SM) algorithm, often referred to as the \textit{relay-type controller}, which is particularly well-suited for systems with relative degree one such as (\ref{CL}). The control input is defined by
\begin{equation}\label{FOSMC}
u(t)=\rho\sign(\sigma(t)) \,,
\end{equation}
where $\rho > 0$ denotes the controller parameter. The right-hand side of the differential equation (\ref{CL}) is discontinuous; therefore, the system trajectories are defined in the sense of a differential inclusion $\dot{\sigma} \in F(\sigma)$. Since $0\in F(0)$, it follows that $\sigma \equiv 0$ is an equilibrium point. A Lyapunov Function (LF) defined as $V(t)=\sigma^2(t)/2$ can be used to demonstrate the finite-time stability of the equilibrium, yielding $\dot{V}\leq-\alpha|\sigma|$ where $\alpha=\rho-L>0$. Consequently, the closed-loop system (\ref{CL}) is insensitive to any uniformly bounded perturbation $|f| \leq L$, where $L > 0$ is known. The necessary and sufficient condition for robust stabilization \cite{Utkin92} is that the control gain satisfies $\rho > L$. The sliding variable converges to zero within a finite time \cite{Shtessel14} bounded by $t_r \leq \sqrt{2V(0)}/\alpha$. \\

\begin{figure}[t]
	\begin{center}
		\vspace{0mm}
		\includegraphics[scale=0.44]{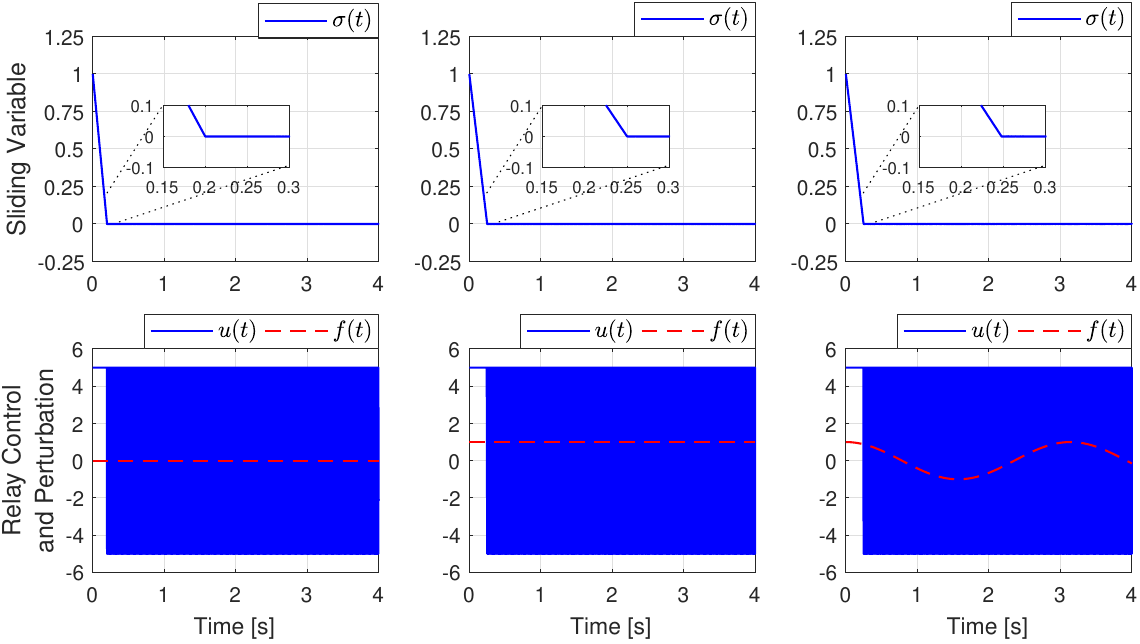}   
	\end{center}
	\vspace{-2mm}
	\caption{Ideal sliding-mode behavior of system (\ref{CL}) with parameter $\rho=5$ and disturbance (\ref{Pertur}):
    \textbf{null} ($\eta=0$) on the left, \textbf{constant} ($\eta=1,\,\Omega=0$) in the center, and \textbf{sinusoidal} ($\eta=1$, $\Omega=2$) on the right.}\label{IdealSM1}
	\vspace{-2mm}
\end{figure}

\textbf{Example 1.} Consider the perturbation signal
\begin{equation}\label{Pertur}
f(t)=\eta \cos\left(\Omega t\right) \,,
\end{equation}
of magnitude $\eta\in\RE$ and some frequency $\Omega\in[0,\hspace{1mm}\Omega_{\text{max}})$. The closed-loop system (\ref{CL}) is solved using Euler’s integration method (ODE1) with a small integration step of $T_s = 10^{-4}$ [s]. The controller parameter is chosen as $\rho = 5$ to set the convergence time  to $t_r \leq 0.25$ [s], starting from the initial condition $\sigma(0) = 1$ and considering the disturbance upper-bound $|f| \leq \eta = 1$. Figure \ref{IdealSM1} shows the time evolution of the error and control signals, considering the disturbance:
\textbf{null} ($\eta=0$) on the left, \textbf{constant} ($\eta=1,\,\Omega=0$) in the center, and \textbf{sinusoidal} ($\eta=1$, $\Omega=2$ [rad/s]) on the right. The following observations can be drawn from the closed-loop behavior:
\begin{itemize}
	\item The sliding variable converges to zero in finite time, with a reaching time bounded by $t_r \leq 0.25$ [s]. No overshoot is observed in the transient response \cite{Utkin20}.
	\item The control input exhibits a discontinuous profile with amplitude $\rho=5$ and a high switching frequency on the order of $T_s^{-1} = 10$ [kHz]. This empirically confirms that the condition $T_s \ll \Omega_{\text{max}}^{-1}$ is satisfied.
	\item First-order sliding mode accuracy is achieved in steady state \cite{LyF10}, with the sliding variable satisfying the bound $|\sigma| \leq \gamma_1 T_s$. The constant $\gamma_1 = 6$ was identified through numerical simulations.
\end{itemize}

\begin{remark}
Several SM algorithms guaranteeing robust stabilization for systems of relative degree one (\ref{CL}) have been extensively studied, including the Sub-optimal \cite{Bartolini98} and the Super-Twisting \cite{Levant98}. More recently, strategies capable of handling systems with arbitrary relative degree under matched disturbances have been proposed, such as the Quasi-Continuous \cite{Levant05a} and the Integral-Discontinuous Control \cite{Moreno20}, among others.
\end{remark}

\subsection{Real Sliding-Modes}
To guarantee exact disturbance rejection and finite-time convergence of the sliding variable, the control signal must ideally switch with infinite frequency \cite{Levant10}. In practice, however, this is unattainable due to non-idealities such as unmodeled actuator and sensor dynamics, time delays, hysteresis, and signal discretization. Modern control systems are commonly implemented on digital platforms optimized for high-speed processing, where the delays associated with analog-to-digital conversion and control law computation are typically negligible compared to the dominant plant time constants. In contrast, the dynamics of input-output transducers are typically fixed—either due to physical constraints or cost considerations. Among these, the actuator often represents the dominant source of delay, not only due to the latency it introduces but also because it imposes a fundamental limit on the achievable bandwidth of the closed-loop system.
\begin{assumption}
    \textit{The frequency response of the actuator–plant cascade dynamics, denoted by $W(j\omega)$, satisfies $|W(j\omega)|\gg|W(j n \omega)|$  for $n = 2, 3, ...$, at the frequency $\omega$ of self-excited oscillations \cite{Boiko04}.}\\
\end{assumption}

\begin{figure}[t]
	\begin{center}
		\vspace{0mm}
		\includegraphics[scale=0.44]{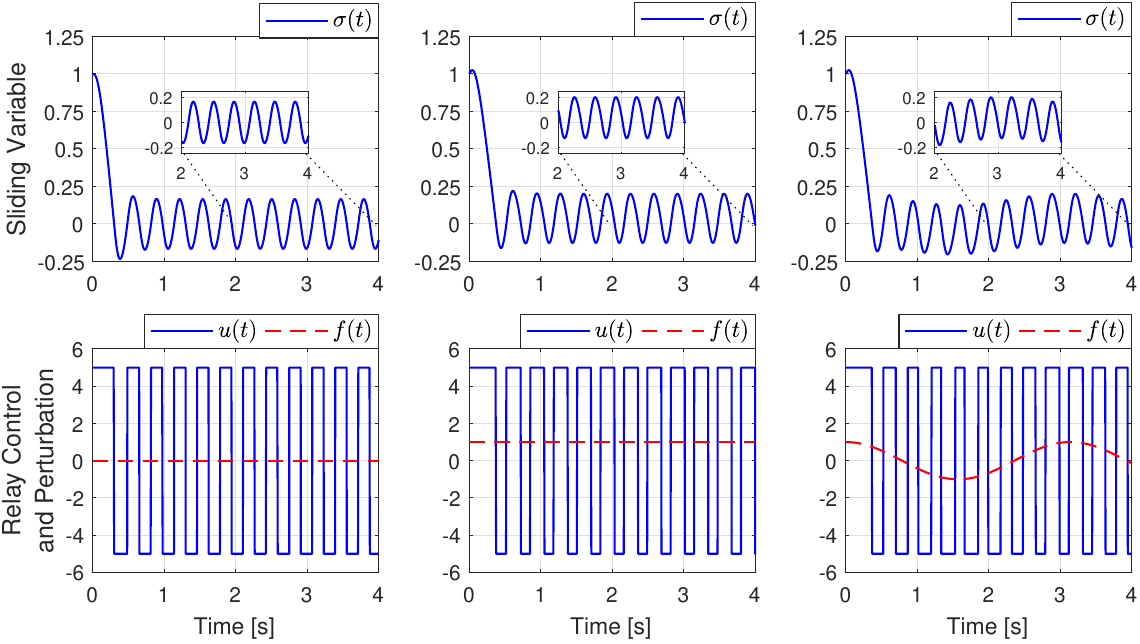}   
	\end{center}
	\vspace{-2mm}
	\caption{Real Sliding-Modes with parameters $\rho=5$, $\mu=0.05$, and disturbance (\ref{Pertur}):
    \textbf{null} ($\eta=0$) on the left, \textbf{constant} ($\eta=1,\,\Omega=0$) in the center, and \textbf{sinusoidal} ($\eta=1$, $\Omega=2$) on the right.}\label{RealSM1}
	\vspace{-2mm}
\end{figure}

\textbf{Example 2.} Consider the actuator dynamics as the primary source of the chattering phenomenon. Following \cite{Utkin20}, the actuator is modeled as a critically damped second-order system
\begin{equation}\label{Actuador}
	\begin{array}{l}
	\dot{z}(t)  =  \begin{bmatrix} 0 & 1 \\ -\frac{1}{\mu^2} & -\frac{2}{\mu} \end{bmatrix}z(t) + \begin{bmatrix} 0 \\ \frac{1}{\mu^2} \end{bmatrix}\bar{u}(t) \,, \\
	u(t)  =  \begin{bmatrix} 1 & 0 \end{bmatrix}  z(t) \,,
	\end{array} 
	\vspace{0mm}
	\end{equation}
where $u(t)$ is the actual actuator output, $\bar{u}(t)$ is the control signal generated by the SM controller, and $z\, \in \, \RE^m$ is the actuator's state vector. The small parameter $0<\mu\ll1$ denotes the actuator time-constant, which is assumed to be known. However, in the design of the \textit{relay-type controller}, the dynamics of the actuator (\ref{Actuador})—with an actuator time-constant of $\mu = 0.05$—are not taken into account. Therefore, the controller parameter is again selected as $\rho = 5$, ensuring the condition $\rho > L$ is satisfied, under the assumption that $|f| \leq \eta = 1$. The initial conditions are set as $z_{1,2}(0) = 0$ for the actuator (\ref{Actuador}) and $\sigma(0) = 1$ for the error dynamics (\ref{CL}). Figure \ref{RealSM1} shows the time evolution of the error and control signals, considering the disturbance:
\textbf{null} ($\eta=0$) on the left, \textbf{constant} ($\eta=1,\,\Omega=0$) in the center, and \textbf{sinusoidal} ($\eta=1$, $\Omega=2$) on the right. The following observations can be made:
\begin{itemize}
\item The sliding variable converges asymptotically to a bounded, oscillatory steady-state regime \cite{Shtessel14}, with a noticeable overshoot observed during the transient response.
\item In the absence of external disturbances ($\eta = 0$), the output exhibits symmetric oscillations centered at the origin, i.e., $\sigma \equiv 0$, as shown in Fig.\ref{RealSM1} (left). In contrast, when disturbances are present ($\eta \neq 0$), the oscillations become asymmetric due to a \textit{bias} introduced by the perturbation \cite{Boiko08}, as illustrated in Fig.\ref{RealSM1} (center and right, corresponding to constant and time-varying disturbances, respectively). 
\end{itemize}
\begin{assumption}
\textit{At steady state, the sliding variable can be approximated by the waveform}
\begin{equation}\label{ModeloSalida}
\sigma(t) = \sigma_0(t) + A\sin(\omega t) \,, \vspace{0mm}
\end{equation}
\textit{under the assumption that the disturbance signal (\ref{Pertur}) evolves slowly compared to the chattering frequency, i.e., $\Omega \ll \omega$.}
\end{assumption}
\begin{remark}
Although the steady‐state sliding variable (\ref{ModeloSalida}) is commonly decomposed into slow‑motions (a distorted version of the perturbation (\ref{Pertur}), attenuated and phase‑shifted) and fast‑motions (chattering effects) \cite{Boiko08, Boiko07}. This interpretation should not be regarded as evidence of a strict superposition principle. Such quasi-linear behavior cannot be expected to hold uniformly across the entire frequency spectrum, nor for arbitrary magnitudes of the external disturbance \cite{Gelb68}.
\end{remark}

\subsection{Problem Statement}
Sliding-mode (SM) control systems implemented with non-ideal actuators often exhibit two-timescale dynamics, characterized by high-frequency fast motions (chattering) and low-frequency slow motions (\textit{bias} drift) in response to external perturbations. Accurately predicting the steady-state behavior of such systems is challenging due to the intrinsic nonlinearity of the control law and the frequency-dependent filtering imposed by actuator dynamics. Classical linearization techniques, particularly small-signal methods, typically fail to capture the multi-scale interaction and nonlinear gain characteristics associated with these behaviors.

To address this, the present work adopts a quasi-linearization framework \cite{Gelb68}, wherein the nonlinear SM control function is approximated by distinct linear operators for fast and slow components of the input. Specifically, the harmonic balance describing function (HB-DF) is used to estimate the amplitude and frequency of fast motions, while an incremental describing function approach is employed to construct frequency response plots (Bode diagrams) of the slow-motion \textit{bias} component. This approach offers a key advantage over conventional small-signal methods: it captures the dependence of the closed-loop response on input amplitude. As a result, the proposed method provides greater insight into the effects of actuator bandwidth and disturbance magnitude on the steady-state performance. Numerical simulations are presented to support the theoretical predictions and to illustrate the range of validity of the quasi-linear approximation.

\section{Preliminaries}
An exact analytical tool for studying the nonlinear phenomena is often more difficult to obtain  than employing an approximate yet sufficiently simple method that captures the system's particular behaviors under specific conditions. Several techniques have been developed to answer different questions about the dynamics of nonlinear systems. The most common approach involves linearization—replacing each nonlinear operation with a linear approximation—and analyzing the resulting system \cite{Khalil92}. However, this technique is only valid in small neighborhoods around an operating point; any system response that extends beyond these limits cannot be accurately captured unless multiple linearizations are performed around different operating points \cite{Apkarian00}. Moreover, many nonlinearities are discontinuous or non-differentiable near the operating point, making conventional linearization inapplicable. To bypass the small-signal restriction while preserving the analytical convenience of linear methods,  an alternative is to characterize the nonlinear element's behavior in response to a \textit{reasonably sized} signal and approximate this behavior with a linear operator. This approach leads to different linear representations for the same nonlinear function, depending on the form or amplitude of the input signal \cite{Atherton75}.

The \textit{quasi-linear} approximation that characterizes the input-output behavior of a nonlinear element under an asymmetric sinusoidal excitation—such as that given by (\ref{ModeloSalida})—is referred to as the Describing Function (DF) \cite{Gelb68}. In the context of SM controllers, two canonical input waveform types are considered \cite{Boiko07,Boiko08}:
\begin{itemize}
    \item[(a)] \textit{\textit{bias}}:  A drift in the input to the nonlinear function is expected. The nonlinearity can produce a biased output even in the absence of a constant input.
    \item[(b)] \textit{Sine wave}:  Any periodic signal generated by the nonlinearity is expected to approximate a \textit{sine} wave after passing through the \textit{low-pass} linear block present in the closed-loop.
\end{itemize}
The structure of the \textit{quasi-linear} approximator comprises a parallel set of linear operators, each corresponding to a distinct \textit{stationary component} of the input signal  \cite{Gelb68}. These operators—often referred to as weighting functions—are designed to minimize the \textit{mean squared error} between the output of the approximator and the actual response of the nonlinearity \cite{Booton54}. A fundamental limitation of this method is its reliance on the input signal closely resembling the assumed waveform (\ref{ModeloSalida}). This constraint introduces a trade-off between the low-pass filtering capability of the linear subsystem and the high-frequency harmonic content generated by the nonlinear SM controller.

\subsection{Sinusoidal-Input Describing Function}
Consider $\sigma_0(t)=\sigma_{0}$ in (\ref{ModeloSalida}), where $\sigma_{0}\in \RE$ is a constant satisfying $|\sigma_{0}|<A$. Then, the input to the SM controller is an \textit{asymmetric sinusoidal} wave, i.e.
\begin{equation}\label{1S_DF}
\sigma(t)=\sigma_{0}+A\sin(\omega t) \,.
\end{equation}
The linear filter that allows the operation of the constant component in (\ref{1S_DF}) is
\begin{equation}\label{BIAS}
\overline{N_{0}}(A,\omega,\sigma_{0})=\dfrac{\omega}{2\pi\sigma_{0}}\int_{0}^{\frac{2\pi}{\omega}}\hspace{0mm}u(t)\,\text{d}t\,, 
\end{equation} 
which is a static gain that depends on the characteristics of the input signal (\ref{1S_DF}). The operator (\ref{BIAS}) captures the average input-output relationship of a given nonlinearity, expressed as
\begin{equation}\label{Cntrl_0}
u^*_0=\overline{N_{0}} \, \sigma_{0} \,. \vspace{0mm}
\end{equation}
The linear filter that allows the operation of the sinusoidal component in (\ref{1S_DF}) is
\begin{equation}\label{DF}
\overline{N_{1}}(A,\omega,\sigma_0) =\overline{N_{P}}
+j\,\overline{N_{Q}} \,,  \vspace{0mm}
\end{equation}
with
\begin{equation*}
\begin{array}{l}
\overline{N_{P}}(A,\omega,\sigma_0)=\dfrac{\omega}{\pi A}\displaystyle\int_{0}^{\frac{2\pi}{\omega}} \hspace{-2mm} u(t)\cdot \sin (\omega t) \, \text{d} t \,,\\
\overline{N_{Q}}(A,\omega,\sigma_0)=\dfrac{\omega}{\pi A}\displaystyle\int_{0}^{\frac{2\pi}{\omega}} \hspace{-2mm} u(t)\cdot \cos (\omega t) \, \text{d} t \,,
\end{array}\vspace{0mm}
\end{equation*}
which is a complex gain in phase and quadrature, determined by the characteristics of the input signal (\ref{1S_DF}). Define the differential operator $s:=\text{d}/\text{d}t$, then the fundamental harmonic component at the output of the nonlinearity can be expressed as
\begin{equation}\label{Cntrl_fund}
u^*_1(t)=\left(\overline{N_{P}}+\frac{\overline{N_{Q}}}{\omega}\, s\right) A\sin(\omega t) \,.
\end{equation} 
\begin{remark}
Any linear filter can be fully characterized by a complex gain that defines how it modifies the amplitude and phase of a sinusoidal input. In contrast, the DF (\ref{DF}) captures the amplitude and phase relationship between the input sinusoid (\ref{1S_DF}) and the fundamental harmonic component (\ref{Cntrl_fund}) at the output of a given nonlinearity \cite{Gelb68}.\\
\end{remark} 

\begin{figure}[t]
	\begin{center}
		\vspace{0mm}
		\includegraphics[scale=0.6]{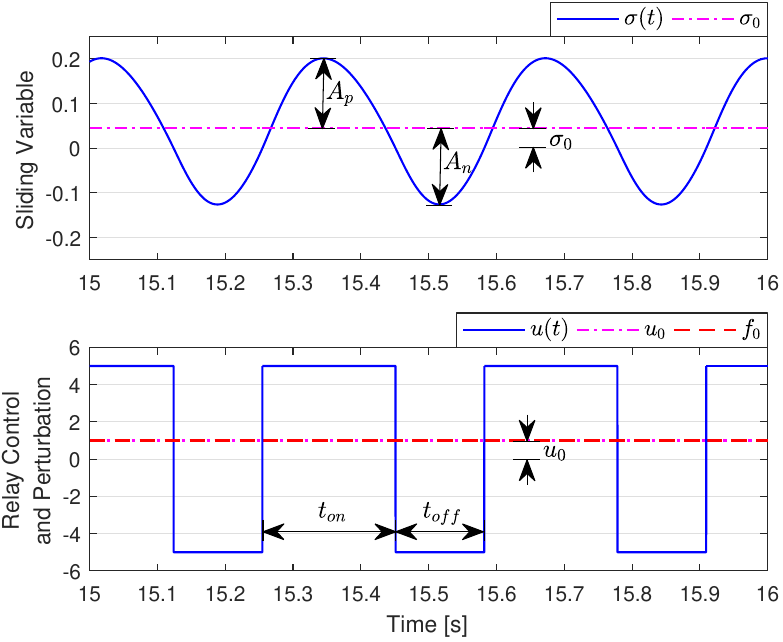}   
	\end{center}
	\vspace{-2mm}
	\caption{Steady-state response with the constant perturbation (\ref{Pertur}) of magnitude $\eta=1$.}\label{Constant}
	\vspace{-2mm}
\end{figure}

Consider the \textit{relay-type controller} operating under an asymmetric sinusoidal input signal as defined in (\ref{1S_DF}). The linear operator (\ref{BIAS}), which enables the propagation of the constant (\textit{bias}) component through this nonlinearity, is given by
\begin{equation}\label{N0_FOSMC}
\overline{N_{0}}(A,\sigma_{0}) =\frac{2\rho}{\pi\sigma_{0}}\,\text{arcsin}\hspace{-0.8mm}\left(\frac{\sigma_{0}}{A}\right)\hspace{-0.5mm} \,. \vspace{0mm}
\end{equation}
The \textit{sign} function is odd which makes $\overline{N_{Q}}=0$ in (\ref{DF}), therefore, the linear filter to operate the sinusoidal component is
\begin{equation}\label{N1_FOSMC}
\overline{N_{1}}(A,\sigma_{0})=\frac{4\rho}{\pi A}\sqrt{1-\left(\frac{\sigma_{0}}{A}\right)^2} \,.\vspace{0mm}
\end{equation}
Figure~\ref{Constant} displays the input and output signals of the \textit{relay-type controller} obtained from the simulations, using  $\rho = 5$, $\mu = 0.05$, and the constant disturbance (\ref{Pertur}) denoted $f_0=\eta$ with magnitude $\eta = 1$. The key chattering parameters observed in the experiment at steady-state are as follows: oscillations amplitude $A=(A_p+A_n)/2=0.1634$, oscillations period $T=t_{\text{on}}+t_{\text{off}}=0.3272$ [s], \textit{bias} component $\sigma_{0}=0.0448$, and average control value $u_{0}=0.9994$. Accordingly, the oscillations frequency is $\omega=\frac{2\pi}{T}=19.2029$ [rad/s], and the ratio $|\sigma_{0}|/A= 0.2742$ is achieved. The average values of the signals over each fast-oscillations period are computed offline and included for reference. By substituting the parameters listed above into expressions (\ref{N0_FOSMC})–(\ref{N1_FOSMC}), the following predictions can be made:
\begin{itemize}
    \item[(i)] The constant component (\ref{Cntrl_0}) at the output of the \textit{relay-type controller} is
    \begin{equation}\label{u0_FOSMC}
    u^*_{0}=0.8840\,, \vspace{0mm}
    \end{equation}
    resulting in a estimation error of $13$\% when compared to the simulation outcome. The average control in the experiment closely converges to $u_0 \approx f_0$ (see Figure~\ref{Constant}, bottom).
    \item[(ii)] The control signal exhibits a square waveform with a duty cycle of $40$\%. The fundamental harmonic (\ref{Cntrl_fund}) at the output of the \textit{relay-type controller} is given by
    \begin{equation*}\label{u1_FOSMC}
    u^*_1(t)=6.122 \sin(19.2029 t) \,. \vspace{0mm}
    \end{equation*}
\end{itemize}

The Sinusoidal-Input Describing Function (SIDF) of the \textit{sign} nonlinearity characterizes the input–output behavior of the \textit{relay-type controller} when driven by an asymmetric sinusoidal input of the form (\ref{1S_DF}), under the condition $|\sigma_0| < A$ \cite{Gelb68}. The prediction error associated with the \textit{bias} component (\ref{u0_FOSMC}) remains within a tolerable margin—less than $15\%$—thus confirming the validity of the DFs (\ref{N0_FOSMC})–(\ref{N1_FOSMC}) in capturing the steady-state response of the nonlinearity (\ref{FOSMC}) under the given conditions.


\subsection{Dual-Input Describing Function}
When a relatively \textit{slow} signal affects the control loop, the steady-state error signal results from a combination of the distorted perturbation—attenuated and phase-shifted—and the high-frequency oscillations  \cite{Boiko08, Boiko07}. Accordingly, a more general input signal for the calculation of the DFs should consist of a single \textit{bias} term together with an arbitrary number of \textit{independent} sinusoidal components \cite{Gelb68}. The input to the nonlinear operator will therefore be considered in its general form as
\begin{equation}\label{2S_DF}
\sigma(t)=\sigma_{0}\cos(\Omega t+ \theta)+A\sin(\omega t) \,, \vspace{0mm}
\end{equation}
that is, a fast component (\textit{chattering}) in addition to a slow component (\textit{\textit{bias}}) resulting from the propagation of a \textit{sinusoidal} perturbation (\ref{Pertur}), such that $|\sigma_{0}| < A$ and $\Omega \ll \omega$. Importantly, when the sinusoidal components in (\ref{2S_DF}) are independent\footnote{Independence is defined in terms of frequency: if the sinusoids were harmonically related, their periods would be proportional, introducing a fixed phase relationship. In contrast, incommensurate frequencies (with an irrational ratio) imply no fixed phase relation, making the relative phase irrelevant to the nonlinear response \cite{Gelb68}.}—that is, their frequencies are incommensurate—the notion of a relative phase angle loses its physical relevance. Under such conditions, it is appropriate to set the phase angle $\theta$ to zero in the quasi-linear approximation.

The linear filter corresponding to each sinusoidal component in (\ref{2S_DF}) is, respectively, a proportional-derivative linear network of the form (\ref{DF}). For \textit{static} and \textit{single-valued} nonlinear functions, the \textit{Dual}-Input Describing Function (DIDF) can be expressed as
\begin{equation}\label{DIDF}
\begin{array}{l}
\widetilde{N_0}(A,\omega,\sigma_{0},\Omega)=\displaystyle\frac{1}{2\pi^2\sigma_{0}}\int\limits_{0}^{2\pi}\text{d}\omega t \int\limits_{0}^{2\pi}u(t)\cdot\sin(\Omega t)\,\text{d}\Omega t  \,,  \\
\widetilde{N_1}(A,\omega,\sigma_{0},\Omega)=\displaystyle\frac{1}{2\pi^2A}\int\limits_{0}^{2\pi}\text{d}\omega t \int\limits_{0}^{2\pi}u(t)\cdot\sin(\omega t)\, \text{d}\Omega t \,, 
\end{array}
\end{equation}
where the integrations could equally well be taken over any other interval which spans one cycle \cite{Gelb68}, such as ($-\pi$, $\pi$). 

\subsubsection{Power-Series Expansion}
A particularly straightforward method for computing the DIDF of odd nonlinearities subjected to non-harmonically related sinusoidal inputs is the power series expansion technique \cite{Gelb68}. The essential steps of this method are briefly summarized below, taking as an example the \textit{relay-type controller} driven by the two-sinusoid input signal defined in (\ref{2S_DF}). Since this is a \textit{static} nonlinearity, the integrals in (\ref{DIDF}) can be expanded as a \textit{Taylor series} around the origin, $\sigma_0 = 0$, as follows
\begin{equation*}
	\begin{array}{l}
	\widetilde{N_0}(A,\sigma_{0})= \displaystyle \dfrac{2\rho}{\pi A}\left(1+\frac{1}{8}\left(\frac{\sigma_{0}}{A}\right)^2+\frac{3}{64}\left(\frac{\sigma_{0}}{A}\right)^4+...\right),\\
	\widetilde{N_1}(A,\sigma_{0})= \displaystyle \dfrac{4\rho}{\pi A}\left(1-\frac{1}{4}\left(\frac{\sigma_{0}}{A}\right)^2-\frac{3}{64}\left(\frac{\sigma_{0}}{A}\right)^4-...\right).
	\end{array} \vspace{0mm}
\end{equation*}
This suggests that the operation of such a nonlinearity on a small signal in the presence of other uncorrelated signals is independent of the form of the small signal \cite{Atherton75}.
\begin{assumption}
\textit{The magnitude of the slow-motion component in the steady-state response (\ref{2S_DF}) is significantly smaller than that of the fast-motion oscillations; that is, $|\sigma_0| \ll A$.}
\end{assumption}

\begin{figure}[t]
	\begin{center}
		\vspace{0mm}
		\includegraphics[scale=0.6]{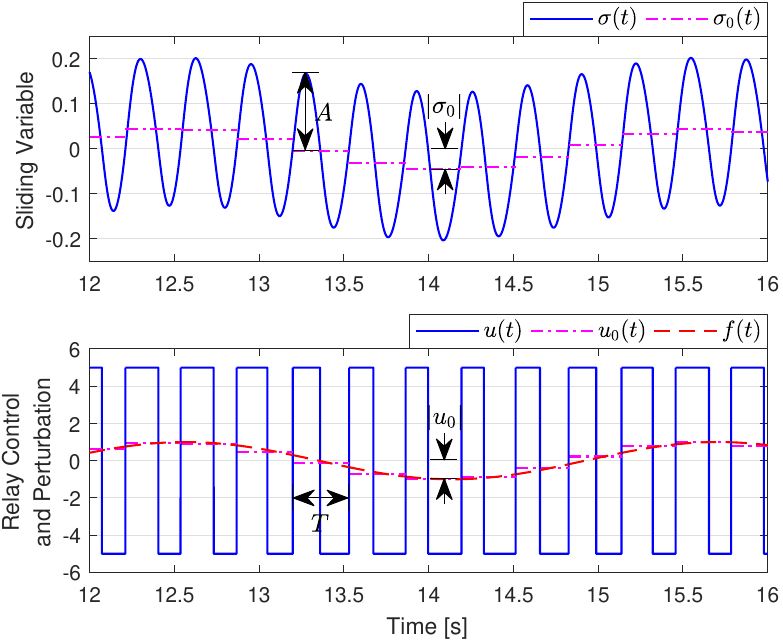}   
	\end{center}
	\vspace{-2mm}
	\caption{Steady-state response with the sinusoidal perturbation  (\ref{Pertur}) of magnitude $\eta=1$ and frequency $\Omega=2$ [rad/s].}\label{Sinusoidal}
	\vspace{-2mm}
\end{figure}

\subsubsection{Incremental-Input Describing Function}
In many practical scenarios, the steady-state response consists of two sinusoidal components, with one exhibiting a significantly smaller amplitude than the other. The \textit{quasi-linear} approximation of a nonlinearity under such a two-sinusoid input—formulated in (\ref{2S_DF}) and consistent with \textit{Assumption~3}—is referred to as the \textit{Incremental}-Input Describing Function (IIDF) \cite{Gelb68}. This local approximation can be derived from the SIDF (\ref{DF}) by computing
\begin{equation}\label{FDEI}
N_i(A,\omega)=\overline{N_{1}}(A,\omega,0)+\frac{A}{2}\, \frac{\text{d}}{\text{d}A}\left\{\overline{N_{1}}(A,\omega,0)\right\} .
\end{equation}
The justification for using the IIDF (\ref{FDEI}) lies in its ability to approximate the results of the integrals in (\ref{DIDF}) near of the origin $\sigma_0 \equiv 0$, as follows: $\widetilde{N_0} \approx N_i(A,\omega)$ and $\widetilde{N_1} \approx \overline{N_{1}}(A,\omega)$, respectively. Then, the fast and slow components at the output of the nonlinearity can be predicted by
\begin{align}
u^*_0(t)=& \overline{N_{i}} \sigma_{0}\cos(\Omega t) \,, \label{Crtl_0s}\\
u^*_1(t)=& \overline{N_{1}} A\sin(\omega t) \,, \label{Crtl_funds}
\end{align}
whenever the conditions $\Omega \ll \omega$ and $|\sigma_{0}| \ll A$ are satisfied. \\

Consider the \textit{relay-type controller} operating under a two-sinusoid input signal as defined in (\ref{2S_DF}). The corresponding SIDF is given in (\ref{N1_FOSMC}), then, a local approximation around $\sigma_{0} = 0$ yields
\begin{equation}\label{FDESe}
\overline{N_{1}}(A)=\frac{4\rho}{\pi A}\left.\sqrt{1-\left(\frac{\sigma_{0}}{A}\right)^2} \right|_{\sigma_{0}=0}=\frac{4\rho}{\pi A} \,. \vspace{0mm}
\end{equation}
Therefore, the IIDF (\ref{FDEI}) of the \textit{relay} takes the form
\begin{equation}\label{FDEIe}
N_i(A)=\dfrac{4\rho}{\pi A}+\frac{A}{2}\, \frac{\text{d}}{\text{d}A}\left\{\dfrac{4\rho}{\pi A}\right\}=\dfrac{2\rho}{\pi A} \,. \vspace{0mm}
\end{equation}
Figure~\ref{Sinusoidal} displays the input and output signals of the \textit{relay-type controller} obtained from the simulations, using  $\rho = 5$, $\mu = 0.05$, and the sinusoidal disturbance (\ref{Pertur}) denoted $f_0=\eta\cos(\Omega t)$ with magnitude $\eta = 1$ and frequency $\Omega=2$ [rad/s]. The key chattering parameters observed in the experiment at steady-state are as follows: fast-oscillations amplitude $A=0.1576$, fast-oscillations period $T=0.332$ [s], upper-bound of \textit{bias} component $|\sigma_{0}|= 0.0447$, and maximum average control value $|u_{0}|=0.9854$. Accordingly, the fast-oscillations frequency is $\omega=18.925$ [rad/s] and the ratios $\Omega/\omega=0.0528$ and $|\sigma_{0}|/A=0.2836$ are achieved. The average values of the signals over each fast-oscillations period are computed offline and included for reference. By substituting the parameters listed above into expressions (\ref{Crtl_0s})–(\ref{Crtl_funds}), the following predictions can be  made:
\begin{itemize}
	\item[(i)] The slow component (\ref{Crtl_0s}) at the output of the \textit{relay-type controller} is
\begin{equation}\label{u0_FOSMC2}
	u^*_{0}(t)=0.9028\cos(t) \,,  \vspace{0mm}
	\end{equation}
	resulting in a estimation error of $9.15$\% when compared to the simulation outcome. The maximum value of the average control closely converges to $|u_0| \approx |f|$ (see Figure \ref{Sinusoidal}, bottom).
	\item[(ii)] The control signal is square with a variable duty cycle; approximately $40$\% when the disturbance reaches its maximum. The fundamental harmonic (\ref{Crtl_funds}) at the output of the \textit{relay-type controller} is given by
	\begin{equation*}\label{u1_FOSMC2}
	u^*_1(t)=6.366 \sin(18.925 t)  \,. \vspace{0mm}
	\end{equation*}
	\end{itemize}

The \textit{Incremental}-Input Describing Function \cite{Gelb68} (IIDF) of the \textit{sign} nonlinearity enables modeling the input-output transfer properties of the \textit{relay-type controller} when driven by an additive sinusoids (\ref{2S_DF}), under the conditions $\Omega\ll\omega$ and $|\sigma_{0}|\ll A$ . The prediction error associated with the \textit{bias} component (\ref{u0_FOSMC2}) remains within a tolerable margin—less than $15\%$—thus confirming the validity of the DFs (\ref{FDESe})–(\ref{FDEIe}) in capturing the steady-state response of the nonlinearity (\ref{FOSMC}) under the given conditions.

\begin{figure}[t]
	\begin{center}
		\vspace{0mm}
		\includegraphics[scale=0.36]{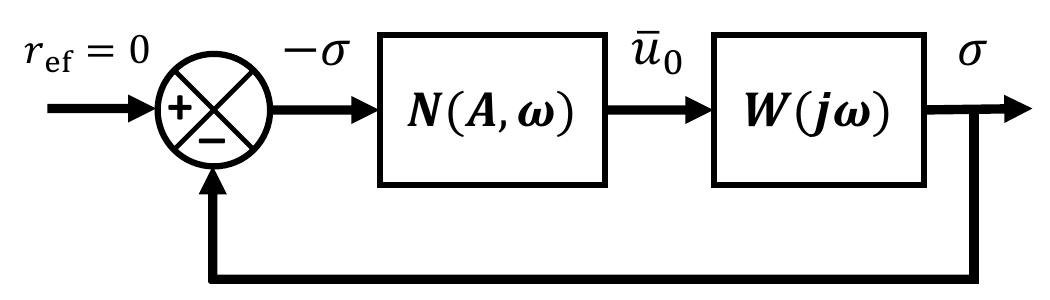}
        \end{center}
	\vspace{-2mm}
	\caption{Quasi-linearized model for the study of fast motions.}\label{Hb_Tesis}
	\vspace{-2mm}
\end{figure}

\subsection{Fast Motions}
To investigate the existence of limit cycles in the undisturbed closed-loop system—i.e., when $f(t) = 0$, implying $\sigma_0(t) = 0$—the SM controller is approximated by its SIDF, $\overline{N_{1}}(A, \omega)$, and the linear subsystem by its frequency response $W(j\omega)$. This yields a \textit{quasi-linearized} of Figure \ref{Hb_Tesis}, wherein the steady-state oscillation $\sigma(t) = A\sin(\omega t)$ leads to the following frequency-domain relation
\begin{equation}\label{RelationSignal}
\begin{bmatrix}
\overline{N_{1}}(A,\omega) & -1 \\
1 & W(j\omega)
\end{bmatrix} \begin{bmatrix}
\sigma(j\omega) \\
\bar{u}(j\omega)
\end{bmatrix} =0 \,. \vspace{0mm}
\end{equation}
The matrix equation (\ref{RelationSignal}) admits non-trivial solutions if and only if its determinant vanishes. This condition is classically reformulated as the harmonic balance (HB) equation,
\begin{equation}\label{EBA}
\overline{N_{1}}(A,\omega)W(j\omega) +1 = 0 \,, \vspace{0mm}
\end{equation}
whose solutions ($A^*,\omega^*$) can represent different response modes, some of which may be unstable \cite{Gelb68}.\\

Consider any linear block  $W(j\omega)$ operating in closed-loop with the \textit{relay-type controller} characterized by the SIDF (\ref{FDESe}). The HB equation (\ref{EBA}) can be generally expressed as
\begin{equation}\label{HB_FOSMC}
\underbrace{\frac{4\rho}{\pi A}+\Real\left\{\frac{1}{W(j\omega)}\right\}}_{\textbf{U}(A,\omega)} + j\underbrace{\Imag\left\{\frac{1}{W(j\omega)}\right\}}_{\textbf{V}(\omega)} =0 \,,  \vspace{0mm}
\end{equation}
whose solution consists of finding
\begin{equation}\label{HB_FOSMC_algo}
\omega^*>0\,:\,\textbf{V}(\omega^*) =0 \hspace{2mm} \Rightarrow \hspace{2mm}
A^*>0 \,:\, \textbf{U}(A^*,\omega^*) =0 \,. \vspace{0mm}
\end{equation}

The linear block is constructed as the cascade connection of the critically damped second-order actuator dynamics, described by (\ref{Actuador}), and the nominal (undisturbed) tracking error dynamics given in (\ref{CL}). The resulting transfer function (TF) is therefore given by
\begin{equation}\label{Linearblock}
W(s)=\dfrac{1}{s(\mu s +1)^2} \,,
\end{equation}
where $0<\mu\ll1$ denotes the actuator time constant. Following the HB-based algorithm (\ref{HB_FOSMC_algo}) applied to the linear block (\ref{Linearblock}), the predicted amplitude and frequency of the main chattering harmonics (\ref{ModeloSalida}) are given by
\begin{equation}\label{Chatt_FOSMC} 
\omega^*=\frac{1}{\mu} \,, \hspace{4mm}
A^*=\frac{2\rho}{\pi}\mu   \,. \vspace{0mm}
\end{equation}
Figure~\ref{RealSM1} (left) shows the input and output signals of the \textit{relay-type controller} obtained from the simulations, using  $\rho = 5$, $\mu = 0.05$, and no disturbance input (\ref{Pertur}), i.e., $\eta = 0$. According to the analytical predictions (\ref{Chatt_FOSMC}), the chattering parameters are $\omega^* = 20$ [rad/s] and $A^* = 0.1591$. In comparison, the simulation measurements yield $\omega = 19.513$ [rad/s] and $A = 0.1655$, resulting in prediction errors of $2.435\%$ for the frequency and $4.023\%$ for the amplitude. 

\subsubsection{Loeb's Stability Criterion}
It is said that a limit cycle is stable if it returns to its original equilibrium state \cite{Gelb68}, while if its amplitude or frequency grows or decays until reaching another equilibrium state, it is considered unstable. Let $(A^*,\omega^*)$ be a solution to the HB equation (\ref{EBA}), that is,
\begin{equation}\label{EBA_Re_Im}
\textbf{U}(A^*,\omega^*) + j \textbf{V}(A^*,\omega^*) = 0 \,, \vspace{0mm}
\end{equation}
where $\textbf{U} = \Real\left\{NW+1\right\}$ and $\textbf{V} = \Imag\left\{NW+1\right\}$. The stability of a predicted \textit{limit cycle} in (\ref{Chatt_FOSMC}), under \textit{quasi-static perturbations}\footnote{Small-deviations in amplitude $A^*=A^*+\Delta A^*$, frequency $\omega^*=\omega^*+\Delta \omega^*$, and amplitude variation $\Delta \nu = - \dot{A}/A$ due to the change in frequency; where by definition the values $\Delta A$, $\Delta\omega$ and $\Delta\nu$ are small.} around the equilibrium $(A^*,\omega^*)$, is characterized by the following inequality
\begin{equation}\label{Estabilidad}
\left.\left(\frac{\partial \textbf{U}}{\partial A} \frac{\partial \textbf{V}}{\partial \omega} - \frac{\partial \textbf{U}}{\partial \omega} \frac{\partial \textbf{V}}{\partial A}\right)\right|_{\substack{A=A^*\\\omega=\omega^*}} > 0  \,. \vspace{0mm}
\end{equation}
This condition, known as Loeb’s criterion \cite{Loeb54}, provides a \textit{necessary} requirement for orbital stability. A more recent study based on the \textit{dynamic harmonic balance} \cite{Boiko18b} (DHB), has introduced a \textit{sufficient} condition for the same purpose. \\

\begin{figure}[t]
	\begin{center}
		\vspace{0mm}
		\includegraphics[scale=0.55]{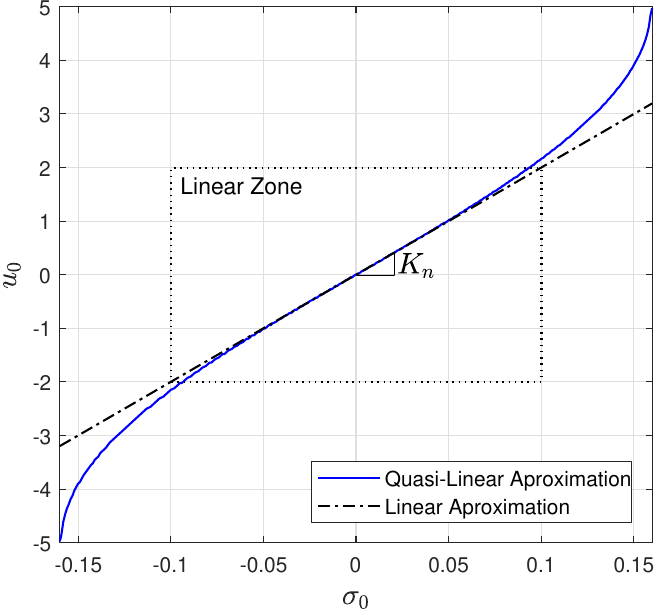}   
	\end{center}
	\vspace{-2mm}
	\caption{Average control function (\ref{u0_FOSMC}) and equivalent gain (\ref{EG_FOSMC}) of the relay-type controller with parameter $\rho = 5$. Given the amplitude $A^*=0.1591$ and the EG $K_n=20$ for the actuator time constant $\mu = 0.05$.}\label{EGain}
	\vspace{-2mm}
\end{figure}

The functions $\textbf{U}(A,\omega)$ and $\textbf{V}(A,\omega)$ that define the HB equation (\ref{HB_FOSMC}) satisfy the following partial derivative relations
\begin{equation*}\label{ParcialMT}
\begin{array}{lcl}
\frac{\partial \textbf{U}}{\partial A} =- \frac{4\rho}{\pi A^2} \,, &\hspace{2mm}& \frac{\partial \textbf{V}}{\partial A} =0 \,,\\
\frac{\partial \textbf{U}}{\partial \omega} = \frac{\Deriv}{\Deriv\omega}\Real\left\{\frac{1}{W(j\omega)}\right\} \,, &\hspace{2mm}& \frac{\partial \textbf{V}}{\partial \omega} = \frac{\Deriv}{\Deriv\omega}\Imag\left\{\frac{1}{W(j\omega)}\right\} \,.\\
\end{array}  
\end{equation*}
Substituting these expressions into the inequality (\ref{Estabilidad}) proposed by \textit{Loeb}, one obtains the \textit{orbital stability condition for relay-type control systems}, 
\begin{equation}\label{LoebMC}
\left. \frac{\Deriv}{\Deriv\omega}\Imag\left\{\frac{1}{W(j\omega)}\right\}\right|_{\substack{\omega=\omega^*}}
<0 \,. \vspace{0mm}
\end{equation} 
For example, consider the linear block (\ref{Linearblock}) for which the \textit{limit cycle} parameters are given by (\ref{Chatt_FOSMC}). The corresponding \textit{orbital stability condition}, as defined by (\ref{LoebMC}), is then
\begin{equation}\label{LoebMC_1}
1-3\mu^2\omega^2\bigg|_{\substack{\omega=\frac{1}{\mu}}}=-2
<0 \,,\vspace{0mm}
\end{equation}
which is satisfied for any small parameter $\mu>0$.

\subsection{Slow Motions}
When the actuation system (\ref{Actuador}) is \textit{static} ($\mu = 0$), the tracking error converges to zero in finite time, driven by the SM controller which \textit{instantaneously} adjusts the control signal in response to disturbances \cite{Utkin92}. In practice, however, the control signal is constrained by the \textit{bandwidth} imposed by actuator dynamics ($\mu > 0$). This limitation gives rise to pulse-width modulation (PWM) behavior \cite{Boiko08}, wherein the control signal consists of a high-frequency carrier component $u_1^*(t) = \overline{N_{1}}(A) \cdot A \sin(\omega t)$ and a low-frequency modulator component $u_0^*(t) = N_{i}(A) \cdot \sigma_0\cos(\Omega t + \phi)$. Demodulation—typically implemented via low-pass filtering of the high-frequency control signal—reveals its average influence on the system. This average action, commonly referred to as the \textit{equivalent control} in the SM literature \cite{Shtessel14}, facilitates the analysis of slow-motion dynamics \cite{Boiko08}. 

The equivalent gain (EG) provides a linear relationship between the average control signal and the \textit{bias} component, i.e., $u_0 = K_n \sigma_0 \approx f_0$. It is defined as the first-order Taylor linearization of the average control function about the origin, where $u_0(0) = 0$. Accordingly, the EG is given by
\begin{equation}\label{EG}
K_n = \left.\dfrac{\partial u_0}{\partial \sigma_0}\right|_{\sigma{0}=0} \,.
\end{equation}
The average control function for the \textit{relay-type controller} was previously calculated using the linear filter (\ref{BIAS}), resulting in (\ref{u0_FOSMC}). Therefore, the EG (\ref{EG}) takes the form 
\begin{equation}\label{EG_FOSMC}
K_n= \left.{\dfrac{2\rho}{\pi A\sqrt{1-\left(\frac{\sigma_{0}}{A}\right)^2}}}\right|_{\sigma_{0}=0}=\dfrac{2\rho}{\pi A} \,. \vspace{0mm}
\end{equation}
The setup of Example 2 is revisited, where the amplitude of fast motions is predicted as $A^* = 2\rho\mu/\pi$, and their frequency is roughly $\omega^* = 1/\mu$. The EG of the \textit{relay-type controller} is then $K_n = 1/\mu$. Figure \ref{EGain} depicts the average control function (\ref{u0_FOSMC}) for a controller parameter of $\rho = 5$, yielding the predicted amplitude of $A^* = 0.1591$ for an actuator time constant $\mu = 0.05$. The graph is evaluated over the \textit{bias} interval $\sigma_0 \in (-A^*,\hspace{1mm}A^*)$. For comparison, the linear approximation $u_0 = K_n \sigma_0$, with $K_n = 20$ as given by (\ref{EG_FOSMC}), is plotted as a straight line through the origin. Thus, the average control $u_0$ exhibits an approximately linear relationship with respect to $\sigma_0$ in a neighborhood of the origin—specifically, for $|\sigma_0| \leq 0.1$. This observation supports the validity of the linear approximation near the origin and confirms the proportional dependence between the average control input and the slow-motion \textit{bias} component under small-signal conditions (cf. \textit{Assumption 3}). 

\begin{figure}[t]
	\begin{center}
		\vspace{0mm}
		\includegraphics[scale=0.36]{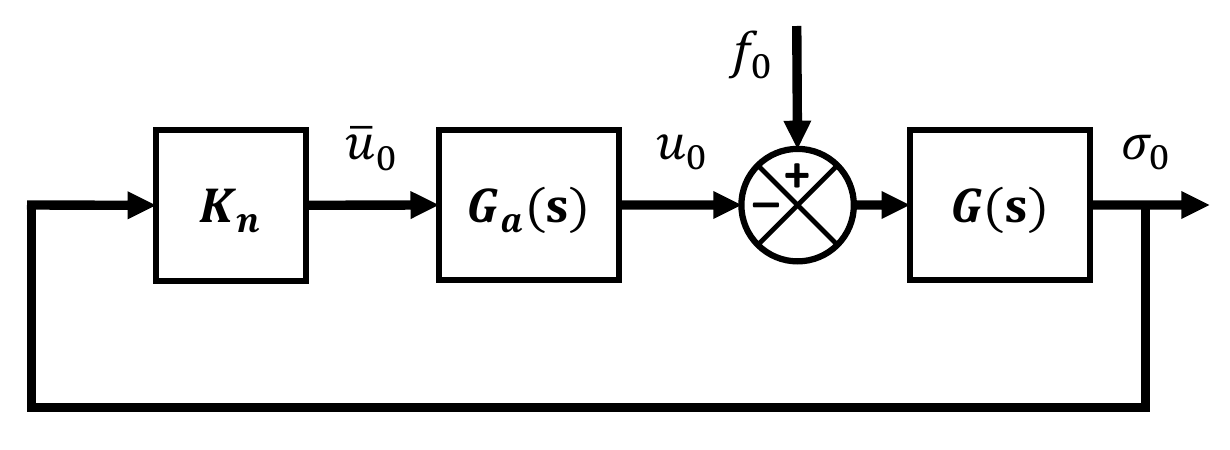}
        \end{center}
	\vspace{-2mm}
	\caption{Linearized model for the study of slow motions.}\label{Kn_CL}
	\vspace{-2mm}
\end{figure}

\begin{figure}[t]
	\begin{center}
		\vspace{0mm}
		\includegraphics[scale=0.6]{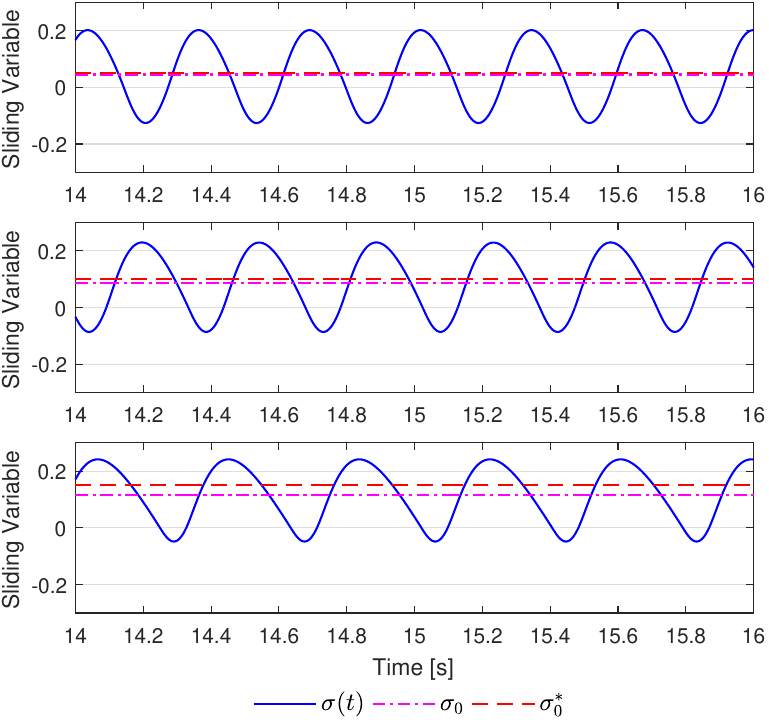}   
	\end{center}
	\vspace{-2mm}
	\caption{Steady-state response under constant perturbations with magnitudes $\eta = 1$, $2$, and $3$ (top to bottom). Theoretical predictions of the \textit{bias} component are given by (\ref{Offset-FOSMC}), respectively.}\label{ConstantAmps}
	\vspace{-2mm}
\end{figure}

\subsubsection{Linearized Model of Propagation}
The EG provides a basis for linearizing the closed-loop dynamics in the presence of slowly varying disturbances and is thus instrumental in constructing analytical propagation models \cite{Boiko08}. Consider a perturbation signal of the form $f(t) = \eta \cos(\Omega t)$. At steady state, the corresponding \textit{bias} component in the sliding variable (\ref{ModeloSalida}) is assumed to take the form $\sigma_{0}(t) = \sigma_{0} \cos(\Omega t + \phi)$. This result is justified by the notion of \textit{equivalent control}, according to which the slow-motions dynamics can be approximated by the cycle-averaged effect of the fast-motions \cite{Fridman01}. Under \textit{Assumptions 2-3} and in accordance with Fig. \ref{Kn_CL}, the \textit{sensitivity transfer function} (see more details in \cite{Boiko07}) allowing to compute the propagation of slow motions is given by
\begin{equation}\label{Sensibilidad}
\frac{\sigma_0(\mathrm{s})}{f_0(\mathrm{s})} = \frac{G(\mathrm{s})}{1 + K_n W(\mathrm{s})} \,, \vspace{0mm}
\end{equation}
where $\mathrm{s} = j\Omega$ denotes the Laplace complex frequency, $G(\mathrm{s})$ is the TF of the plant, and $W(\mathrm{s})=G_a(\mathrm{s})G(\mathrm{s})$ corresponds to the overall actuator–plant cascade. The parameter $K_n$ is the EG of the relay-type controller computed in (\ref{EG}).

\begin{table}[b]
	\centering
	\scalebox{1.4}{
		\begin{tabular}{|c||c|c|c|}
			\hline
			\multicolumn{1}{|c||}{} & {\tiny $\eta=1$} & {\tiny $\eta=2$} & {\tiny $\eta=3$} \\ \hline \hline 
			{\tiny $\sigma^*_0$} & \begin{tiny} 0.05 \end{tiny} & \begin{tiny} 0.1 \end{tiny} & \begin{tiny} 0.15 \end{tiny} \\ \cline{1-4} {\tiny $\sigma_0$} & \begin{tiny} 0.0448 \end{tiny} & \begin{tiny} 0.0854 \end{tiny} & \begin{tiny} 0.1146 \end{tiny}\\  \cline{1-4} 
			{\tiny  \textbf{Error}} & \begin{tiny} 10.40\% \end{tiny} & \begin{tiny} 14.60\% \end{tiny} & \begin{tiny} 23.60\% \end{tiny} \\ \hline 
	\end{tabular}}
	\vspace{2mm}
	\caption{\textit{Bias} component in the steady-state response under constant perturbations with magnitudes $\eta = 1$, $2$, and $3$, respectively.}\label{ResultsConstant}	
	\vspace{-2mm}
\end{table}

\section{Bode Plots for Discontinuous Control Design}
It is important to emphasize that the superposition principle does not hold for the sensitivity TF (\ref{Sensibilidad}); in particular, the response to a composite disturbance cannot be constructed as the sum of responses to individual sinusoidal components \cite{Gelb68}. Furthermore, the validity of the propagation model (\ref{Sensibilidad}) depends critically on the assumption that the average control function (\ref{Cntrl_0}) operates within its linear region—namely, the domain in which the equivalent gain (\ref{EG}) provides an accurate local approximation. 

\subsection{Slow Motions due to Constant Perturbations} 
Let the exogenous input be defined as $f_0(\mathrm{s}) = \frac{\eta}{\mathrm{s}}$, where the magnitude satisfies $|\eta| \ll AK_n$. The corresponding \textit{bias} level in the steady-state response (\ref{1S_DF}) can be computed from the sensitivity TF (\ref{Sensibilidad}) as
\begin{equation}\label{Offset}
	\sigma^*_{0}= \lim\limits_{\mathrm{s}\to0} \mathrm{s}\hspace{0.3mm}\sigma_{0}(\mathrm{s})\Bigg\lvert_{f_0(\mathrm{s})=\frac{\eta}{\mathrm{s}}} = \lim\limits_{\mathrm{s}\to0}\frac{\eta\, G(\mathrm{s})}{1 + K_n W(\mathrm{s})} \,.\vspace{2mm}
\end{equation}

\begin{figure}[t]
	\begin{center}
		\vspace{0mm}
		\includegraphics[scale=0.55]{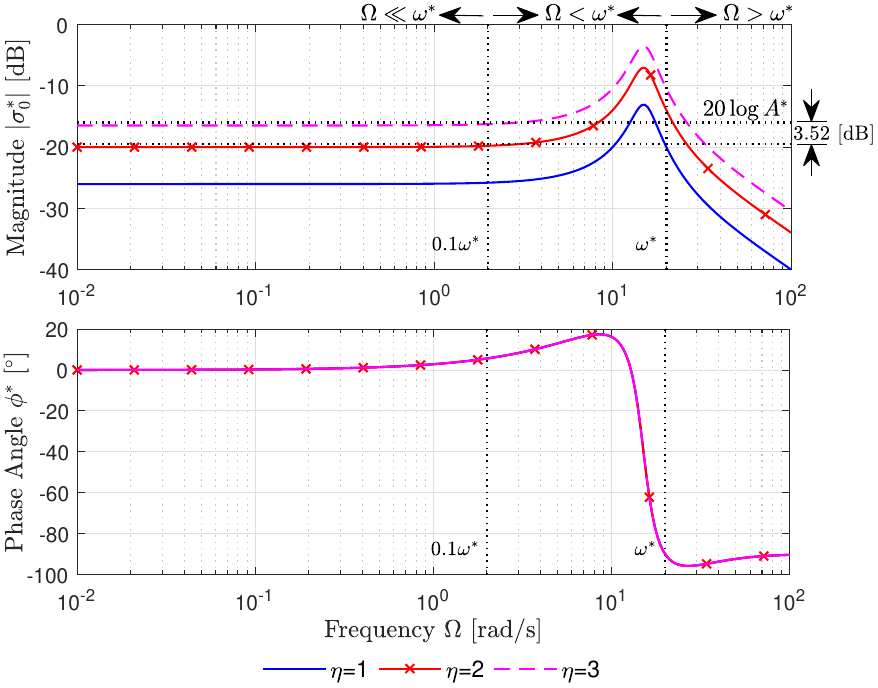}   
	\end{center}
	\vspace{-2mm}
	\caption{Bode plots of the sensitivity transfer function (\ref{Sensibilidad}) over the frequency range $\Omega \in (0.01,\hspace{1mm} 100)$ [rad/s], for perturbation magnitudes $\eta = 1$, $2$, and $3$, respectively.}\label{BodeSign}
	\vspace{-2mm}
\end{figure}

Once the EG given in (\ref{EG_FOSMC}) has been determined, the \textit{bias} component in the sliding variable (\ref{1S_DF}) can be computed using the expression (\ref{Offset}), namely
\begin{equation}\label{Offset-FOSMC}
	\sigma^*_{0}= \lim\limits_{\mathrm{s}\to0}\frac{\eta\, G(\mathrm{s})}{1 + K_n W(\mathrm{s})} = \eta\mu \,,\vspace{0mm}
\end{equation}
where $G(\mathrm{s})=1/\mathrm{s}$, $G_a(\mathrm{s})=1/(\mu \mathrm{s}+1)^2$, and $K_n=1/\mu$. Further simulations are carried out under the same conditions as those described in Example~2, namely with controller gain $\rho = 5$ and actuator time constant $\mu = 0.05$. However, distinct magnitudes of the constant disturbance (\ref{Pertur}) are considered: $\eta = 1$, $2$, and $3$—illustrated from top to bottom in Figure~\ref{ConstantAmps}. A comparison between the theoretical predictions, obtained from (\ref{Offset-FOSMC}), and the corresponding simulation outcomes is provided in Table~\ref{ResultsConstant}. The estimation error of the \textit{bias} component increases with the magnitude of the disturbance. For disturbance levels up to $|\eta| = 2$, the approximation remains sufficiently accurate, with an error below 15\%—provided that \textit{Assumption~2-3} holds. Beyond this threshold, not only the estimations of \textit{bias} level (\ref{Offset-FOSMC}) exhibit significant deviation, but the shape of the sliding variable signal also becomes noticeably distorted.



\begin{remark}
The average control function (\ref{u0_FOSMC}) associated with the \textit{relay-type controller} is fundamentally governed by an \textit{arc-sine} function, which exhibits an approximately linear behavior—with a relative error below 10\% between both theoretically predicted curves shown in Fig.~\ref{EGain}—provided that the ratio between the \textit{bias} magnitude and the carrier amplitude satisfies $|\sigma_0| / A < 2/3$. Accordingly, for a given upper bound $L > 0$ on the constant perturbation (\ref{Pertur}), the controller parameter $\rho > 0$ must be selected to satisfy the following conditions:
\begin{itemize}
    \item $\rho > \eta$ guarantees exact disturbance rejection under ideal sliding-mode conditions ($\mu = 0$), as discussed in Section~I-A.
    \item $\rho > 1.57\eta$ ensures the validity of the quasi-linearized model $u_0^* = \frac{2\rho}{\pi} \, \text{arcsin}\left(\frac{\sigma_{0}^*}{A^*}\right)$—derived from the linear filter (\ref{BIAS})—under practical sliding-mode conditions ($\mu > 0$), as outlined in Section~II-A.
    \item $\rho > 2.36\eta$ justifies the use of the linear approximation $u_0^* = K_n \sigma_0^*$—based on the EG  (\ref{EG_FOSMC})—for predicting the \textit{bias} component (\ref{Offset-FOSMC}) under practical sliding-mode conditions ($\mu > 0$).
\end{itemize}
\end{remark}


\subsection{Slow Motions due to Sinusoidal Perturbations} 
Let the exogenous input be defined as $f(\mathrm{s}) = \frac{\eta\mathrm{s}}{\mathrm{s}^2 + \Omega^2}$, where the disturbance magnitude satisfies $|\eta| \ll AK_n$, and the excitation frequency is assumed to lie in the low-frequency range, i.e., $\Omega \ll \omega$. Under these conditions, the magnitude $|\sigma_0|$ [dB] and phase angle $\phi$ [$^\circ$] of the \textit{bias} component in (\ref{2S_DF}) become frequency-dependent, as they are shaped by the frequency response of the linear subsystem (\ref{Linearblock}) \cite{Boiko07}.\\

The setup of Example 2 is revisited, where the amplitude of fast motions is predicted as $A^* = 2\rho\mu/\pi$, and their frequency is roughly $\omega^* = 1/\mu$. The EG of the \textit{relay-type controller} is then $K_n = 1/\mu$. Figure~\ref{BodeSign} displays the Bode plots of the sensitivity TF (\ref{Sensibilidad}) across frequencies $\Omega \in (0.01,\hspace{1mm}100)$~[rad/s], using $\rho = 5$, $\mu = 0.05$, and several magnitudes of the sinusoidal perturbation (\ref{Pertur}), specifically $\eta = 1$, $2$, and $3$.  Some reference lines are included in both the magnitude and phase plots of Fig.~\ref{BodeSign} to facilitate the interpretation of the frequency response characteristics: 

\subsubsection{Vertical lines}
The sinusoidal disturbance (\ref{Pertur}) may be considered slow-varying when its frequency satisfies $\Omega \ll \omega^*$, consistent with \textit{Assumption 2}. Accordingly, the frequency axis can be qualitatively segmented into three frequency bands:
\begin{itemize}
\item \textbf{Low-Frequency} ($\Omega < 0.1\omega^*$): The system behaves similarly to the constant perturbation case; the \textit{bias} level is about $|\sigma_0^*| \approx \eta\mu$ [dB] and phase angle near to $\phi^* \approx 0$ [$^\circ$].
\item \textbf{High-Frequency} ($0.1\omega^* < \Omega < \omega^*$): The assumptions underlying the \textit{averaging approach} are not strictly satisfied \cite{Fridman01}, and consequently, the EG-based linearized model (\ref{Sensibilidad}) is no longer applicable.
\item \textbf{Cutoff-Frequency} ($\Omega > \omega^*$): Consistent with \textit{Assumption~1}, the actuator dynamics (\ref{Actuador}) rise to a \textit{stop band} in the system's frequency response characterized by an attenuation slope about  $-40$ [dB/decade].
\end{itemize}

\begin{figure}[t]
	\begin{center}
		\vspace{0mm}
		\includegraphics[scale=0.6]{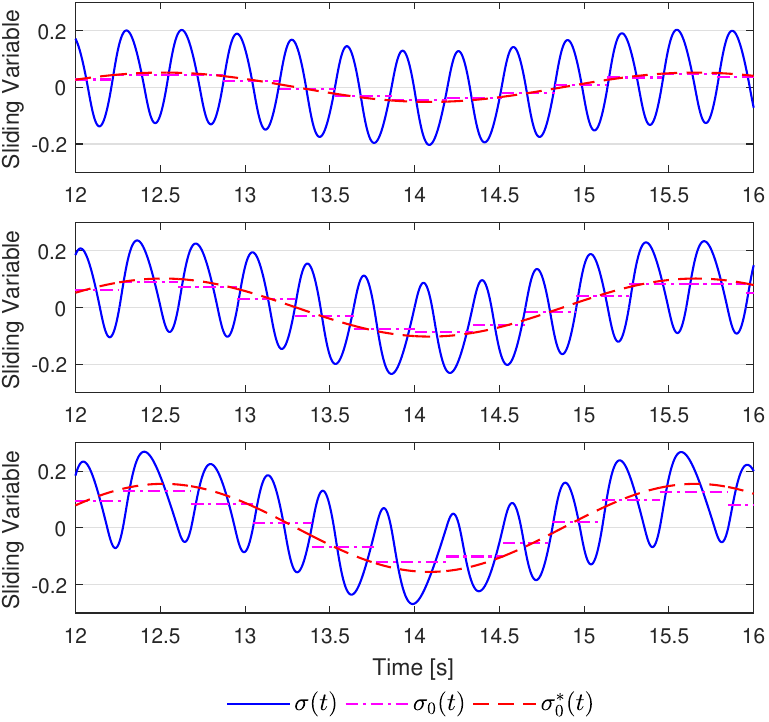}
        \end{center}
	\vspace{-2mm}
	\caption{Steady-state response under sinusoidal perturbations with frequency $\Omega = 2$ [rad/s] and magnitudes $\eta = 1$, $2$, and $3$ (top to bottom). Theoretical predictions of the \textit{bias} component are given by (\ref{VarOffset-FOSMC}), resp.}\label{BodeSignOutputs}
	\vspace{-2mm}
\end{figure}

\begin{table}[b]
	\centering
	\scalebox{1.4}{
		\begin{tabular}{|c||c|c|c|}
			\hline
			\multicolumn{1}{|c||}{} & {\tiny $\eta=1$} & {\tiny $\eta=2$} & {\tiny $\eta=3$} \\ \hline \hline 
			{\tiny $|\sigma^*_0|$} & \begin{tiny} 0.0513 \end{tiny} & \begin{tiny} 0.1026 \end{tiny} & \begin{tiny} 0.1538 \end{tiny} \\ \cline{1-4} {\tiny $|\sigma_0|$} & \begin{tiny} 0.0454 \end{tiny} & \begin{tiny} 0.0892 \end{tiny} & \begin{tiny} 0.1292 \end{tiny}\\  \cline{1-4} 
			{\tiny  \textbf{Error}} & \begin{tiny} 11.50\% \end{tiny} & \begin{tiny} 13.06\% \end{tiny} & \begin{tiny} 15.99\% \end{tiny} \\ 
			\hline 
	\end{tabular}}
	\vspace{2mm}
	\caption{\textit{Bias} component in the steady-state response under sinusoidal perturbations with frequency $\Omega=2$ [rad/s] and magnitudes $\eta = 1$, $2$, and $3$, respectively.}\label{ResultsSinusoidal}	
	\vspace{-2mm}
\end{table}

\subsubsection{Horizontal line (magnitude plot only)} 
Since the \textit{bias} magnitude $|\sigma_0^*|$ depends on the perturbation frequency $\Omega$, the applicability of the linearized propagation model (\ref{Sensibilidad})—which relies on the EG defined in (\ref{EG_FOSMC})—is restricted to frequency intervals where \textit{Assumption 3} remains valid.
\begin{itemize}
\item The line at $20\log A^*$ [dB] corresponds to the predicted amplitude of the fast chattering motions.
\item The line at $20\log A^* - 3.52$ [dB] indicates the threshold associated with the condition $|\sigma_0^*| < \frac{2}{3}A^*$, which ensures the applicability of the EG-based model:
\begin{itemize}
\item For $\eta = 1$ $\Rightarrow$ $\Omega_{\text{max}} = 10.36$ [rad/s].
\item For $\eta = 2$ $\Rightarrow$ $\Omega_{\text{max}} = 3.05$ [rad/s].
\item For $\eta = 3$, the condition is not fulfilled over the entire \textbf{Low-Frequency} band.
\end{itemize}
\end{itemize}
Further simulations are carried out under the same conditions as those described in Example~2, namely with controller gain $\rho = 5$, actuator time constant $\mu = 0.05$, and perturbation frequency $\Omega = 2$ [rad/s]. However, distinct magnitudes of the sinusoidal disturbance (\ref{Pertur}) are considered: $\eta = 1$, $2$, and $3$—illustrated from top to bottom in Figure~\ref{BodeSignOutputs}. The theoretical predictions, obtained by evaluating the sensitivity TF (\ref{Sensibilidad}) at the perturbation frequency $\Omega = 2$ [rad/s], are summarized as follows:
\begin{equation}\label{VarOffset-FOSMC}
    \begin{array}{ccc}
        \text{For} \hspace{1.5mm} \eta=1 & \Rightarrow  \sigma_0^*(t) =0.0513\cos(2t+5.654^\circ). \\
        \text{For} \hspace{1.5mm} \eta=2 & \Rightarrow  \sigma_0^*(t) =0.1026\cos(2t+5.654^\circ). \\
        \text{For} \hspace{1.5mm} \eta=3 & \Rightarrow  \sigma_0^*(t) =0.1538\cos(2t+5.654^\circ). \\
    \end{array}
\end{equation}
A comparison within the theoretical predictions (\ref{VarOffset-FOSMC}) and the corresponding simulation outcomes is provided in Table~\ref{ResultsSinusoidal}. The estimation error of the \textit{bias} component increases with the magnitude of the disturbance. For disturbance magnitudes up to $|\eta| = 2$, the approximation remains sufficiently accurate, with an error below 15\%—provided that \textit{Assumption~2-3} holds. Although precise quantification of the phase prediction error is challenging, a qualitative analysis suggests that for disturbance magnitudes $\eta = 1$ and $\eta = 2$, the bias phases are accurately captured, as shown in Fig.~\ref{BodeSignOutputs} (top and center). In contrast, for $\eta = 3$ (bottom), a noticeable discrepancy arises between the cycle-averaged simulation data and the theoretically reconstructions by means of the sensitivity TF (\ref{Sensibilidad}).

\begin{remark}
Focusing on the magnitude plot in Fig.~\ref{BodeSign} (top), the validity domains of \textit{Assumptions 2-3} can be visualized as the rectangular region enclosed by the coordinate axes, the vertical line at $0.1\omega^*$ [rad/s], and the horizontal line at $20\log(A^*) - 3.52$ [dB]. Within this region, the  sensitivity TF (\ref{Sensibilidad}) based on the EG (\ref{EG_FOSMC}) remains applicable. 
\end{remark}

\begin{figure}[t]
	\begin{center}
		\vspace{0mm}
		\includegraphics[scale=0.55]{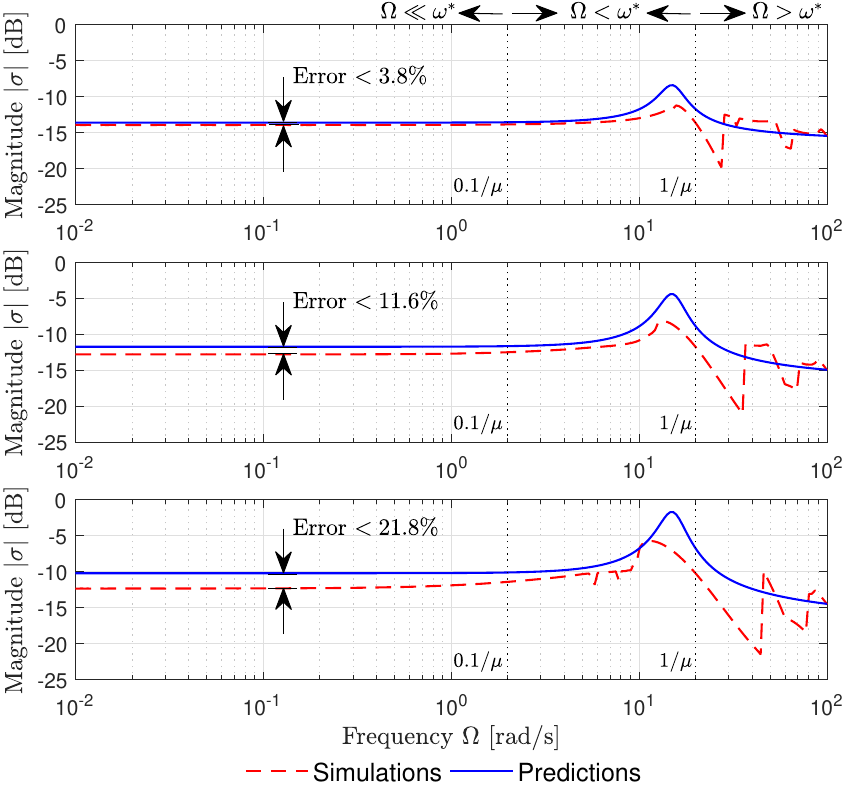}   
	\end{center}
	\vspace{-2mm}
	\caption{Total deviation plots for the sinusoidal disturbance (\ref{Pertur}) of frequency $\Omega\in(0.01,\hspace{1mm} 100)$ [rad/s] and magnitudes $\eta=1$, $2$ and $3$. Theoretical predictions of the \textit{bias} component are given by (\ref{Sensibilidad}), respectively.}\label{BodeSignTotal}
	\vspace{-2mm}
\end{figure}

\subsection{Total Deviation}
The \textit{total deviation} of the sliding variable at steady state can be defined as
\begin{equation}\label{Total_Dev}
	|\sigma|:=\max\limits_{t_r\ll t\leq \text{T}}|\sigma(t)| \,, \hspace{4mm} \text{T} = 2\pi \left(\frac{1}{\Omega}+\mu\right)  \,,
\end{equation}
where $\text{T}$ accounts for both the excitation period associated with the perturbation frequency $\Omega$, as well as the period of the self-excited oscillations corresponding to the characteristic frequency $\omega^* \approx 1/\mu$ induced by the actuator dynamics. Within the rectangular region delineated in \textit{Remark~7}—bounded by $0.1\omega^*$ in frequency and $20\log(A^*) - 3.52$ [dB] in magnitude—the coexistence of fast and slow motion components is well justified. In this domain, the application of the DFs (\ref{FDESe})–(\ref{FDEIe}) to model both components remains valid. Accordingly, the \textit{total deviation} (\ref{Total_Dev}) of the sliding variable can be conservatively estimated as $|\sigma^*| \leq |\sigma_0^*| + A^*$, where $|\sigma_0^*|$ represents the magnitude of slow motions, as predicted by the sensitivity TF (\ref{Sensibilidad}), and $A^*$ denotes the amplitude of fast motions given by (\ref{Chatt_FOSMC}). \\

Figure~\ref{BodeSignTotal} presents the total deviation of the sliding variable at steady state, as computed from simulations using the definition in (\ref{Total_Dev}), over the frequency range $\Omega \in (0.01,\hspace{1mm}100)$~[rad/s]. The simulations were carried out with controller parameters $\rho = 5$, $\mu = 0.05$, and three different magnitudes of the sinusoidal perturbation (\ref{Pertur}), namely $\eta = 1$, $2$, and $3$. The corresponding theoretical predictions are obtained by evaluating the magnitude of the sensitivity transfer function (\ref{Sensibilidad})—with equivalent gain $K_n = \omega^* = 1/\mu$—and summing point-wise the amplitude of the self-excited oscillations, given by $A^* = 2\rho\mu/\pi$. The prediction errors shown in Fig.~\ref{BodeSignTotal} are valid exclusively within the \textbf{Low-Frequency} band ($\Omega < 0.1/\mu$), where the \textit{Assumptions 2-3} remain justified. The plots corresponding to the \textbf{High-Frequency} ($0.1/\mu < \Omega <  1/\mu$) and \textbf{Cutoff-Frequency} ($\Omega > 1/\mu$) regions are included solely for qualitative illustration. 



\section{Bode Plots for Lipschitz Control Design}
The control design strategy considered here involves shaping the control signal through its time derivative. In this formulation, the control input is obtained as the integral of a high-frequency switching function, resulting in a continuous control signal despite the discontinuity in its derivative \cite{Shtessel14}. Let the tracking error be denoted by $\sigma \in \mathbb{R}$, whose dynamics are governed by the first-order system of equation (\ref{CL}). The sliding variable is now defined as
\begin{equation}\label{SlidingVs}
    S(t) = \dot{\sigma}(t) + b\sigma(t) \,.
\end{equation}
where $b > 0$ is a design parameter. Implementation of the so-called \textit{Lipschitz continuous} sliding-mode controller requires differentiation of the error signal and takes the form
\begin{equation}\label{LipschitzCont}
    \dot{u}(t)=\rho\sign(S(t)) \,,
\end{equation}
with $\rho > 0$ as the controller parameter. Finite-time convergence of the sliding variable (\ref{SlidingVs}) to the origin can be demonstrated using the Lyapunov function $V(t) = S^2(t)/2$, which yields the inequality $\dot{V} \leq -\alpha|\sigma|$, where $\alpha = \rho - \bar{L} - bL > 0$. Here, the perturbation $f(t)$ and its derivative are assumed to be uniformly bounded, satisfying $|f| \leq L$ and $|\dot{f}| \leq \bar{L}$. Thus, once the zero dynamics ($S\equiv0$) is achieved in finite time $t\geq t_r$, the exponential convergence of tracking error satisfy $\sigma(t)=\sigma(0)\exp(-bt)$. The necessary and sufficient condition for robust stabilization is that the controller gain satisfies $\rho > \bar{L} + bL$ \cite{Utkin92}. Under this condition, the convergence time is bounded by $t_r \leq \sqrt{2V(0)}/\alpha$. It is important to note that the initial value of the sliding variable $S(0)$ in (\ref{SlidingVs}) depends not only on the initial tracking error $\sigma(0)$, but also on the instantaneous values of the perturbation signal $f(0)$ and its derivative $\dot{f}(0)$. \\




\begin{figure}[t]
	\begin{center}
		\vspace{0mm}
		\includegraphics[scale=0.35]{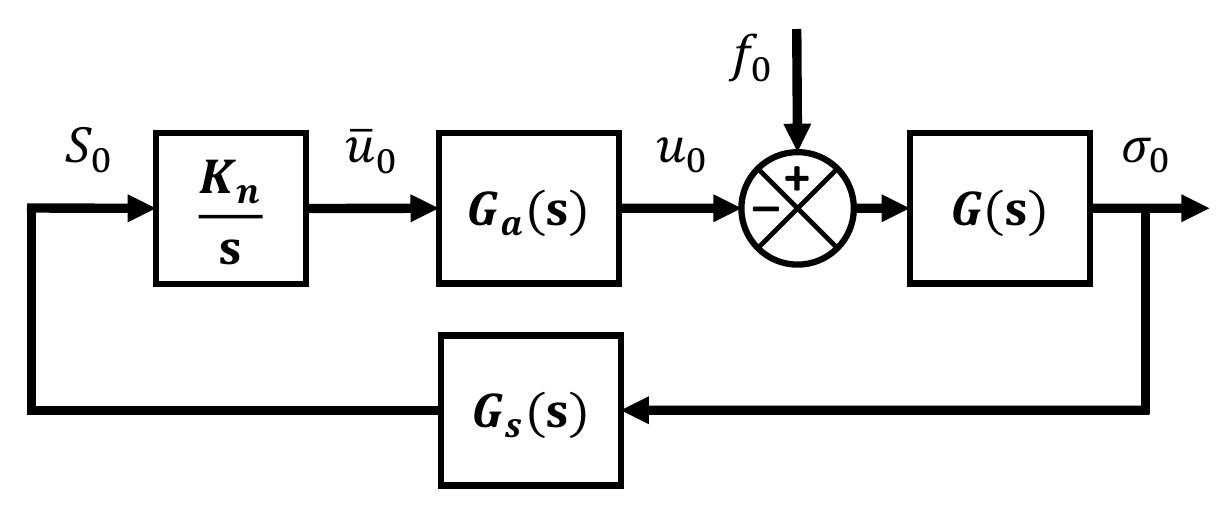}
        \end{center}
	\vspace{-2mm}
	\caption{Linearized model for the study of slow motions.}\label{Kn_Lips}
	\vspace{-2mm}
\end{figure}

According to \cite{Martinez2021}, the linear block is constructed as the cascade of the critically damped second-order actuator dynamics (\ref{Actuador}), the nominal (undisturbed) tracking error dynamics (\ref{CL}), and the sensor-based reconstruct TF is given by
\begin{equation}\label{Linearblock2}
W(s)=\dfrac{s+b}{s^2(\mu s +1)^2} \,,
\end{equation}
where $0<\mu\ll1$ denotes the actuator time constant. By applying the HB-based algorithm (\ref{HB_FOSMC_algo}) to the linear block (\ref{Linearblock2}), the amplitude and frequency of the main chattering harmonics in (\ref{ModeloSalida}) are predicted as
\begin{equation}\label{Chatt_Lipschitz}
\omega^*=\dfrac{\sqrt{1-2b\mu}}{\mu}\,, \hspace{4mm} A^*=\dfrac{2\rho\mu}{\pi\left(1-2b\mu\right)} \,,
\end{equation}
\begin{remark}
The chattering parameters predicted by (\ref{Chatt_Lipschitz}) correspond to the modified sliding variable defined in (\ref{SlidingVs}). The associated amplitude of oscillations in the original tracking error signal can be computed as
\begin{equation}
    a^*=\dfrac{2\rho\mu^2}{\pi(1-2\mu b)(1-\mu b)} \,,
\end{equation}
while the frequency of fast motions remains unchanged.
\end{remark}

\subsubsection{Loeb's Stability Criterion}
According to \cite{Martinez2021}, the corresponding orbital stability condition (\ref{LoebMC}) for the linear block defined in (\ref{Linearblock2}), reduces to
\begin{equation}\label{LoebLips}
1-2b\mu>0 \,,
\end{equation}
the design parameter must satisfy $b < 1/(2\mu)$ for a given actuator time constant $\mu$ to ensure orbital stability \cite{Martinez2021}.

\subsubsection{Linearized Models of Propagation}
The EG of the \textit{relay-type controller} is (\ref{EG_FOSMC}). Then, substituting the amplitude of fast-motions given by (\ref{Chatt_Lipschitz}), the EG particularly results $K_n=\sqrt{1-2b\mu}/\mu$. Unlike in the case of discontinuous sliding modes, the \textit{bias} component here appears simultaneously in both the sliding variable and the tracking error. Under \textit{Assumptions 2-3} and in accordance with Fig. \ref{Kn_Lips}, the \textit{sensitivity transfer functions} allowing to compute the propagation of slow motions are given by:
\begin{itemize}
    \item Sensitivity of the sliding variable to slow perturbations.
\begin{equation}\label{Sensibilidads}
\frac{S_0(\mathrm{s})}{f_0(\mathrm{s})} = \frac{G(\mathrm{s})G_s(\mathrm{s})}{1 + \frac{K_n}{\mathrm{s}} W(\mathrm{s})} \,, \vspace{0mm}
\end{equation}
    \item Sensitivity of the tracking error to slow perturbations.
\begin{equation}\label{Sensibilidadx}
\frac{\sigma_0(\mathrm{s})}{f_0(\mathrm{s})} = \frac{G(\mathrm{s})}{1 + \frac{K_n}{\mathrm{s}} W(\mathrm{s})} \,, \vspace{0mm}
\end{equation}
\end{itemize}
where $\mathrm{s} = j\Omega$ denotes the Laplace complex frequency, $G(\mathrm{s})$ is TF of the plant, $G_s(\mathrm{s})$ represents the sensor dynamics, and $W(\mathrm{s}) = G_a(\mathrm{s})G(\mathrm{s})G_s(\mathrm{s})$ corresponds to the overall actuator–plant–sensor cascade. The parameter $K_n$ is the EG of the relay-type controller computed in (\ref{EG}).

\begin{figure}[t]
	\begin{center}
		\vspace{0mm}
		\includegraphics[scale=0.65]{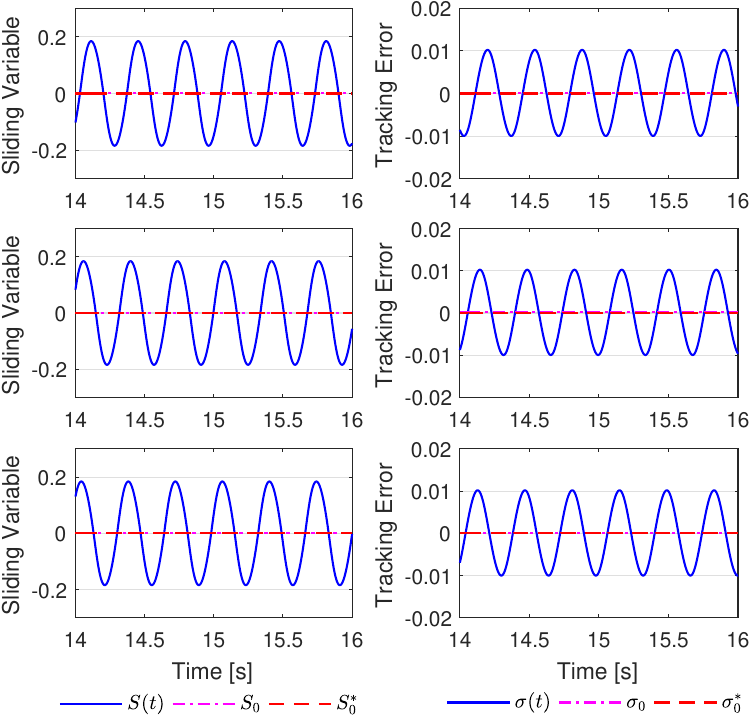} 
	\end{center}
	\vspace{-2mm}
	\caption{Steady-state response under constant perturbations  with magnitudes $\eta = 1$, $2$, and $3$ (top to bottom). Theoretical predictions of the \textit{bias} component are given by (\ref{Offset-FOSMCs})-(\ref{Offset-FOSMCx}), respectively.}\label{ConstantAmpsSup}
	\vspace{-2mm}
\end{figure}

\subsection{Slow Motions due to Constant Perturbations}
Once the EG given in (\ref{EG_FOSMC}) has been determined, the \textit{bias} component in the sliding variable can be computed using the expression (\ref{Sensibilidads}), namely
\begin{equation}\label{Offset-FOSMCs}
	S^*_{0}= \lim\limits_{\mathrm{s}\to0}\frac{\eta\, G(\mathrm{s})G_s(\mathrm{s})}{1 + \frac{K_n}{\mathrm{s}} W(\mathrm{s})} = 0 \,,
\end{equation}
where $G(\mathrm{s})=1/\mathrm{s}$, $G_a(\mathrm{s})=1/(\mu \mathrm{s}+1)^2$, $G_s(\mathrm{s})=\mathrm{s}+b$, and $K_n=(1-2b\mu)/\mu$. On the other hand, the \textit{bias} level in the tracking error can be computed from the sensitivity TF (\ref{Sensibilidadx}), resulting
\begin{equation}\label{Offset-FOSMCx}
	\sigma^*_{0}= \lim\limits_{\mathrm{s}\to0}\frac{\eta\, G(\mathrm{s})}{1 + \frac{K_n}{\mathrm{s}} W(\mathrm{s})} = 0 \,,
\end{equation}
The constant perturbation (\ref{Pertur}) satisfy $|\dot{f}|\leq\bar{L}=0$ and $|f|\leq L=\eta$, uniformly in time. This means that the controller parameter in (\ref{LipschitzCont}) must be selected $\rho>b\eta$ in \textit{ideal} conditions ($\mu=0$). Several simulations are conducted using the parameters $b = 1$,  $\rho = 5$, $\mu = 0.05$, and various magnitudes of the constant disturbance (\ref{Pertur}), specifically $\eta = 1$, $2$, and $3$, as depicted from top to bottom in Figure~\ref{ConstantAmpsSup}. At first glance, the steady-state responses align well with the theoretical predictions (\ref{Offset-FOSMCs})–(\ref{Offset-FOSMCx}); notably, the \textit{bias} levels in both the sliding variable and the tracking error remain effectively negligible. According to the analytical predictions (\ref{Chatt_FOSMC}), the chattering parameters are $\omega^* = 18.974$ [rad/s], $A^* = 0.1768$ and $a^*=0.0093$. In comparison, the simulation measurements yield $\omega = 18.479$ [rad/s], $A = 0.1841$, and $a=0.0101$, resulting in prediction errors of $2.609\%$ for the frequency, $4.129\%$ for the amplitude of the sliding variable and $8.602\%$ for the amplitude of the tracking error.    

\begin{figure}[t]
	\begin{center}
		\vspace{0mm}
		\includegraphics[scale=0.55]{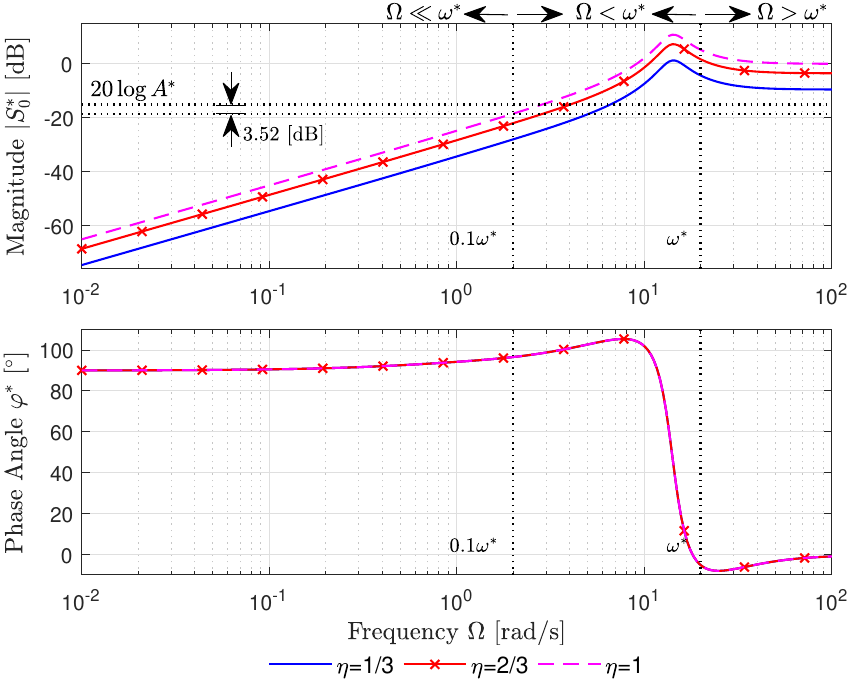}   
	\end{center}
	\vspace{-2mm}
	\caption{Bode plots of the sensitivity transfer function (\ref{Sensibilidads}) over the frequency range $\Omega \in (0.01,\hspace{1mm} 100)$ [rad/s], for perturbation magnitudes $\eta = 1/3$, $2/3$, and $1$, respectively.}\label{BodeSignSups}
	\vspace{-2mm}
\end{figure}

\begin{figure}[t]
	\begin{center}
		\vspace{0mm}
		\includegraphics[scale=0.55]{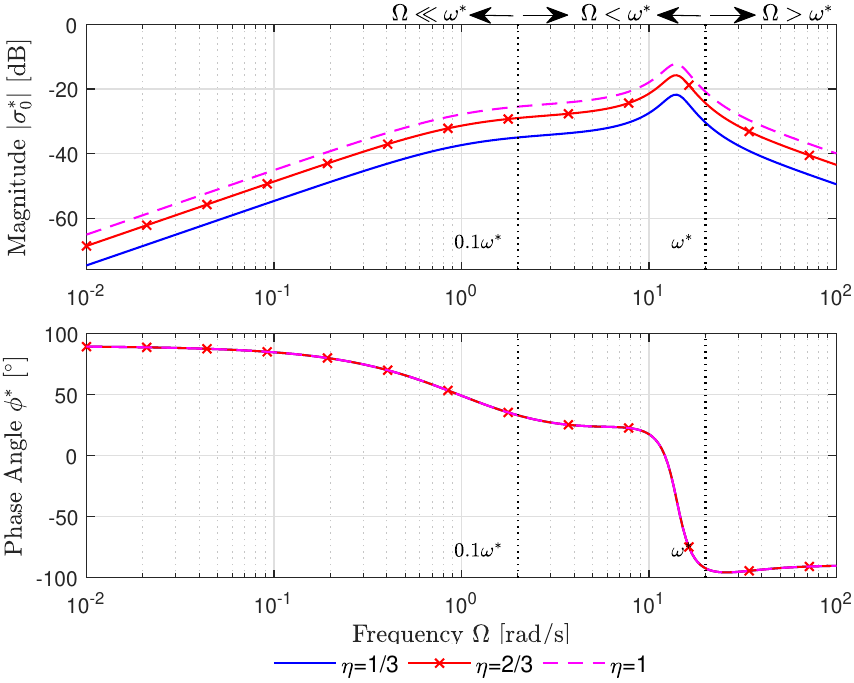}   
	\end{center}
	\vspace{-2mm}
	\caption{Bode plots of the sensitivity transfer function (\ref{Sensibilidadx}) over the frequency range $\Omega \in (0.01,\hspace{1mm} 100)$ [rad/s], for perturbation magnitudes $\eta = 1/3$, $2/3$, and $1$, respectively.}\label{BodeSignSupx}
	\vspace{-2mm}
\end{figure}

\begin{table}[b]
	\centering
	\scalebox{1.4}{
		\begin{tabular}{|c||c|c|c|}
			\hline
			\multicolumn{1}{|c||}{} & {\tiny $\eta=1/3$} & {\tiny $\eta=2/3$} & {\tiny $\eta=1$} \\ \hline \hline 
			{\tiny $|S^*_0|$} & \begin{tiny} 0.0397 \end{tiny} & \begin{tiny} 0.0794 \end{tiny} & \begin{tiny} 0.1191 \end{tiny} \\ \cline{1-4} {\tiny $|S_0|$} & \begin{tiny} 0.0349 \end{tiny} & \begin{tiny} 0.0696 \end{tiny} & \begin{tiny} 0.1029 \end{tiny}\\  \cline{1-4} 
			{\tiny  \textbf{Error}} & \begin{tiny} 12.09\% \end{tiny} & \begin{tiny} 12.34\% \end{tiny} & \begin{tiny} 13.60\% \end{tiny} \\ \hline 
            \hline
            {\tiny $|\sigma^*_0|$} & \begin{tiny} 0.0178 \end{tiny} & \begin{tiny} 0.0355 \end{tiny} & \begin{tiny} 0.0533 \end{tiny} \\ \cline{1-4} {\tiny $|\sigma_0|$} & \begin{tiny} 0.0154 \end{tiny} & \begin{tiny} 0.0302 \end{tiny} & \begin{tiny} 0.0443 \end{tiny}\\  \cline{1-4} 
			{\tiny  \textbf{Error}} & \begin{tiny} 13.48\% \end{tiny} & \begin{tiny} 14.93\% \end{tiny} & \begin{tiny} 16.88\% \end{tiny} \\ \hline
	\end{tabular}}
	\vspace{2mm}
	\caption{\textit{Bias} component in the steady-state response under the sinusoidal perturbation (\ref{Pertur}) with frequency $\Omega=2$ [rad/s] and magnitudes $\eta = 1/3$, $2/3$, and $1$, respectively.}\label{ResultsSinusoidalOut}	
	\vspace{-2mm}
\end{table}

\begin{figure}[t]
	\begin{center}
		\vspace{0mm}
		\includegraphics[scale=0.6]{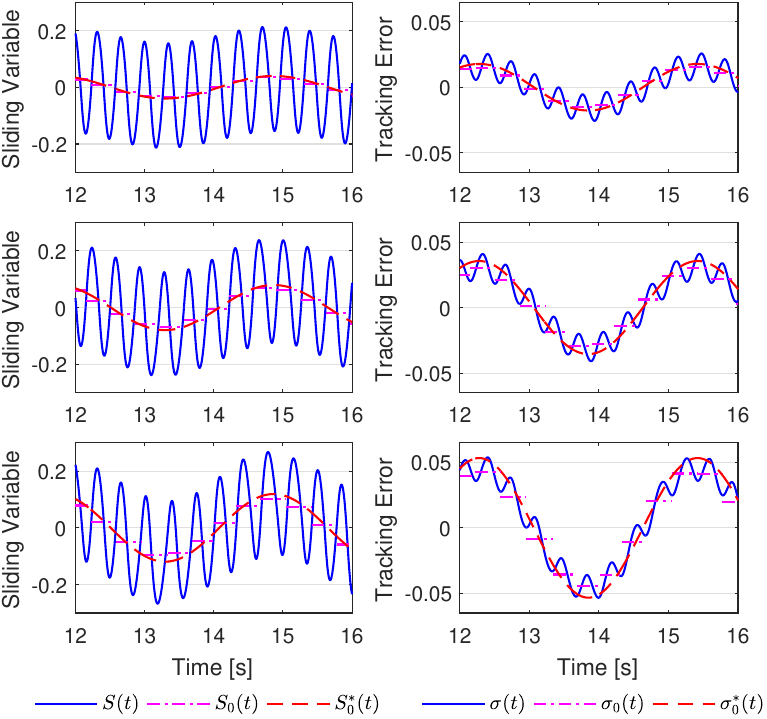}
        \end{center}
	\vspace{-2mm}
	\caption{Steady-state response under sinusoidal perturbations with frequency $\Omega = 2$ [rad/s] and magnitudes $\eta = 1/3$, $2/3$, and $1$ (top to bottom). Theoretical predictions of the \textit{bias} component are given by (\ref{VarOffset-FOSMCs})-(\ref{VarOffset-FOSMCx}), resp.}\label{BodeSignSupOutputs}
	\vspace{-2mm}
\end{figure}

\begin{figure}[t]
	\begin{center}
		\vspace{0mm}
		\includegraphics[scale=0.55]{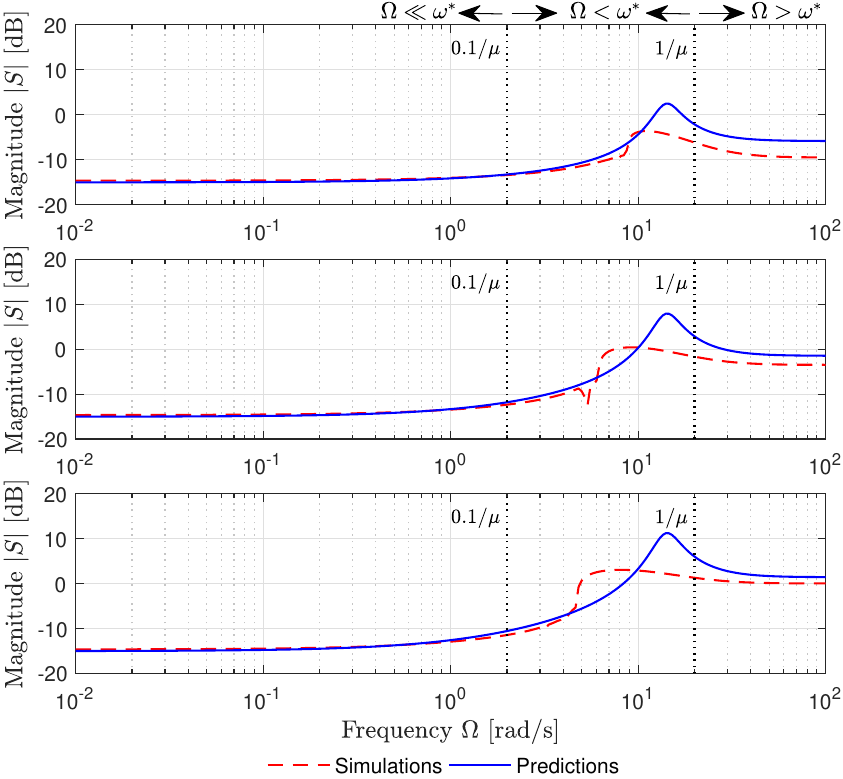}   
	\end{center}
	\vspace{-2mm}
	\caption{Total deviation plots for the sinusoidal disturbance (\ref{Pertur}) of frequency $\Omega\in(0.01,\hspace{1mm} 100)$ [rad/s] and magnitudes $\eta=1$, $2$ and $3$. Theoretical predictions of the \textit{bias} component are given by (\ref{Sensibilidads}), respectively.}\label{BodeSignTotals}
	\vspace{-2mm}
\end{figure}

\begin{figure}[t]
	\begin{center}
		\vspace{0mm}
		\includegraphics[scale=0.55]{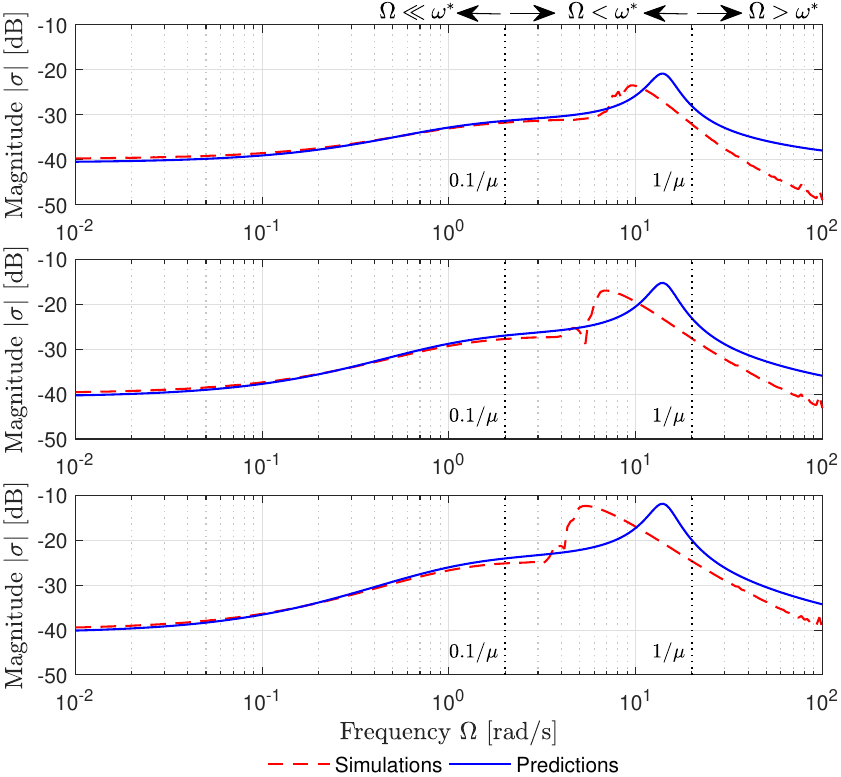}   
	\end{center}
	\vspace{-2mm}
	\caption{Total deviation plots for the sinusoidal disturbance (\ref{Pertur}) of frequency $\Omega\in(0.01,\hspace{1mm} 100)$ [rad/s] and magnitudes $\eta=1$, $2$ and $3$. Theoretical predictions of the \textit{bias} component are given by (\ref{Sensibilidadx}), respectively.}\label{BodeSignTotalx}
	\vspace{-2mm}
\end{figure}

\subsection{Slow Motions due to Sinusoidal Perturbations}
The sinusoidal perturbation (\ref{Pertur}) satisfy $|\dot{f}|\leq\bar{L}=\eta\Omega$ and $|f|\leq L=\eta$, uniformly in time. This means that the controller parameter in (\ref{LipschitzCont}) must be selected $\rho>\eta(\Omega+b)$ in \textit{ideal} conditions ($\mu=0$). Further simulations are performed using the parameters $b = 1$, $\rho = 5$, and $\mu = 0.05$, under sinusoidal disturbances (\ref{Pertur}) of frequency $\Omega = 2$ [rad/s] and varying magnitudes $\eta = 1/3$, $2/3$, and $1$, illustrated from top to bottom in Figure \ref{BodeSignSupOutputs}. The theoretical steady-state responses of the sliding variable are obtained by evaluating the sensitivity TF (\ref{Sensibilidads}) at the perturbation frequency, yielding:
\begin{equation}\label{VarOffset-FOSMCs}
    \begin{array}{lcc}
        \text{For} \hspace{1.5mm} \eta=1/3 & \Rightarrow S_0^*(t) =0.0397\cos(2t+96.6^\circ). \\
        \text{For} \hspace{1.5mm} \eta=2/3 & \Rightarrow S_0^*(t) =0.0794\cos(2t+96.6^\circ). \\
        \text{For} \hspace{1.5mm} \eta=1 & \Rightarrow S_0^*(t) =0.1191\cos(2t+96.6^\circ). \\
    \end{array}
\end{equation}
Similarly, the predicted steady-state responses of the tracking error, derived from the sensitivity TF (\ref{Sensibilidadx}), are given by:
\begin{equation}\label{VarOffset-FOSMCx}
    \begin{array}{lccc}
        \text{For} \hspace{1.5mm} \eta=1/3 & \Rightarrow \sigma_0^*(t) =0.0178\cos(2t+33.17^\circ). \\
        \text{For} \hspace{1.5mm} \eta=2/3 & \Rightarrow \sigma_0^*(t) =0.0355\cos(2t+33.17^\circ). \\
        \text{For} \hspace{1.5mm} \eta=1 & \Rightarrow \sigma_0^*(t) =0.0533\cos(2t+33.17^\circ). \\
    \end{array}
\end{equation}
A comparison between the theoretical predictions in (\ref{VarOffset-FOSMCs})–(\ref{VarOffset-FOSMCx}) and the corresponding simulation results is summarized in Table~\ref{ResultsSinusoidalOut}. Although precise quantification of the phase prediction error is inherently challenging, a qualitative assessment indicates that for disturbance magnitudes $\eta = 1$ and $\eta = 2$, the predicted phase shifts closely match the simulation results, as illustrated in Fig.~\ref{BodeSignSupOutputs} (top and center). However, for the largest disturbance magnitude, $\eta = 3$ (bottom), a clear mismatch emerges between the cycle-averaged simulation response and the theoretical reconstructions obtained via the sensitivity TFs (\ref{Sensibilidads})–(\ref{Sensibilidadx}).

\subsection{Total Deviation}
Figure~\ref{BodeSignTotal} presents the total deviation of the sliding variable at steady state, as computed from simulations using the definition in (\ref{Total_Dev}), over the frequency range $\Omega \in (0.01,\hspace{1mm}100)$~[rad/s]. The corresponding theoretical predictions are obtained by evaluating the magnitude of the sensitivity transfer function (\ref{Sensibilidad})—with equivalent gain $K_n = \omega^* = \sqrt{1-2b\mu}/\mu$—and summing point-wise the amplitude of the self-excited oscillations, given by $A^* = 2\rho\mu/(\pi\sqrt{1-2b\mu})$. The predictions are valid exclusively within the \textbf{Low-Frequency} band ($\Omega < 0.1/\mu$), where the \textit{Assumptions 2-3} remain justified. The plots corresponding to the \textbf{High-Frequency} ($0.1/\mu < \Omega <  1/\mu$) and \textbf{Cutoff-Frequency} ($\Omega > 1/\mu$) regions are included solely for qualitative illustration.

\section{Conclusion}
This work presents a unified frequency-domain framework for the analysis and design of sliding-mode control systems, applicable to both discontinuous and Lipschitz-continuous implementations. Closed-form expressions are derived for the amplitude and frequency of chattering oscillations, along with equivalent gain (EG) models that enable closed-loop sensitivity analysis. The theoretical results reveal the interplay between actuator bandwidth, controller gains, and external disturbances.

For both constant and sinusoidal perturbations, the proposed sensitivity transfer function approach accurately predicts the steady-state bias and oscillatory response within the low-frequency regime ($\Omega < 0.1\omega^*$). Simulation results corroborate the theoretical predictions, showing strong agreement in chattering amplitude, frequency, and phase characteristics. Notably, the bias components of the sliding variable and tracking error remain within a 15\% error margin, , where the EG approximation remains valid, i.e. $|\sigma_0^*| < \frac{2}{3}A^*$ is satisfied. 

The study highlights the importance of respecting actuator/sensor constraints in the selection of controller gains, as formalized by Loeb’s stability condition. The proposed framework offers a principled methodology for robustness analysis in sliding-mode control, facilitating systematic tuning of controller parameters to achieve a trade-off between disturbance rejection and chattering attenuation.

\section*{Acknowledgment}
The authors are grateful for the financial support of CONAHCYT
   (Consejo Nacional de Humanidades, Ciencias y Tecnologías): CVU 631266.

\bibliographystyle{ieeetr}
\bibliography{main.bib}  

\begin{thebibliography}{10}

\bibitem{Ogata02}
K.~Ogata, {\em Modern control engineering, 4th ed.}
\newblock New Jersey: Prentice Hall, 2002.

\bibitem{Utkin92}
V.~Utkin, {\em Sliding modes in optimization and control problems}.
\newblock New York: Springer Verlag, 1992.

\bibitem{Shtessel14}
Y.~Shtessel, C.~Edwards, L.~Fridman, and A.~Levant, {\em Sliding mode control and observation}.
\newblock Boston: Birkh\"{a}user, 2014.

\bibitem{Emelianov86}
S.~Emelianov, S.~Korovin, and L.~Levantovskii, ``Higher-order sliding modes in binary control systems,'' {\em Soviet Physics Doklady}, vol.~31, no.~4, pp.~291--293, 1986.

\bibitem{Bartolini98}
G.~Bartolini, A.~Ferrara, and E.~Usai, ``Chattering avoidance by second-order sliding mode control,'' {\em IEEE Transactions on Automatic Control}, vol.~43, no.~2, pp.~241--246, 1998.

\bibitem{Feng02}
Y.~Feng, X.~Yu, and Z.~Man, ``Non-singular terminal sliding mode control of rigid manipulators,'' {\em Automatica}, vol.~38, no.~12, pp.~2159--2167, 2002.

\bibitem{Levant05a}
A.~Levant, ``Quasi-continuous high-order sliding-mode controllers,'' {\em IEEE Transactions on Automatic Control}, vol.~50, no.~11, pp.~1812--1816, 2005.

\bibitem{Barbot09}
J.-P. Barbot, D.~Boutat, and T.~Floquet, ``An observation algorithm for nonlinear systems with unknown inputs,'' {\em Automatica}, vol.~45, no.~8, pp.~1970--1974, 2009.

\bibitem{Rios21}
H.~R{\'\i}os, R.~Franco, A.~F. de~Loza, and D.~Efimov, ``A high-order sliding-mode adaptive observer for uncertain nonlinear systems,'' {\em IEEE Transactions on Automatic Control}, 2021.

\bibitem{Levant98}
A.~Levant, ``Robust exact differentiation via sliding mode technique,'' {\em automatica}, vol.~34, no.~3, pp.~379--384, 1998.

\bibitem{Angulo12}
M.~T. Angulo, J.~A. Moreno, and L.~Fridman, ``The differentiation error of noisy signals using the generalized super-twisting differentiator,'' in {\em 2012 IEEE 51st Annual Conference on Decision and Control (CDC)}, (Maui, HI, USA), pp.~7383--7388, IEEE, 2012.
\newblock \url{https://doi.org/10.1109/CDC.2012.6426662}.

\bibitem{Chalanga13}
A.~Chalanga, S.~Kamal, and B.~Bandyopadhyay, ``Continuous integral sliding mode control: A chattering free approach,'' in {\em 2013 IEEE International Symposium on Industrial Electronics (ISIE)}, (Taipei, Taiwan), pp.~1--6, IEEE, 2013.
\newblock \url{https://doi.org/10.1109/ISIE.2013.6563756}.

\bibitem{Fridman15}
L.~Fridman, J.~A. Moreno, B.~Bandyopadhyay, S.~Kamal, and A.~Chalanga, ``Continuous nested algorithms: The fifth generation of sliding mode controllers,'' {\em Recent Advances in Sliding Modes: From Control to Intelligent Mechatronics}, vol.~24, pp.~5--35, 2015.

\bibitem{Ding16}
S.~Ding, A.~Levant, and S.~Li, ``Simple homogeneous sliding-mode controller,'' {\em Automatica}, vol.~67, pp.~22--32, 2016.

\bibitem{Laghrouche17}
S.~Laghrouche, M.~Harmouche, and Y.~Chitour, ``Higher order super-twisting for perturbed chains of integrators,'' {\em IEEE Transactions on Automatic Control}, vol.~62, pp.~3588 -- 3593, 2017.

\bibitem{Cruz17}
E.~Cruz-Zavala and J.~A. Moreno, ``Homogeneous high order sliding mode design: A lyapunov approach,'' {\em Automatica}, vol.~80, pp.~232--238, 2017.

\bibitem{Moreno20}
J.~A. Moreno, E.~Cruz-Zavala, and {\'A}.~Mercado-Uribe, ``Discontinuous integral control for systems with arbitrary relative degree,'' in {\em Variable-Structure Systems and Sliding-Mode Control}, pp.~29--69, Springer, 2020.

\bibitem{Panathula17}
C.~B. Panathula, A.~Rosales, Y.~B. Shtessel, and L.~M. Fridman, ``Closing gaps for aircraft attitude higher order sliding mode control certification via practical stability margins identification,'' {\em IEEE Transactions on Control Systems Technology}, vol.~26, no.~6, pp.~2020--2034, 2017.

\bibitem{Levant10}
A.~Levant, ``Chattering analysis,'' {\em IEEE Transactions on Automatic Control}, vol.~55, no.~6, pp.~1380--1389, 2010.

\bibitem{Shtessel96}
Y.~B. Shtessel and Y.-J. Lee, ``New approach to chattering analysis in systems with sliding modes,'' in {\em Proceedings of 35th IEEE Conference on Decision and Control}, vol.~4, (Kobe, Japan), pp.~4014--4019, IEEE, 1996.
\newblock \url{https://doi.org/10.1109/CDC.1996.577355}.

\bibitem{Castillo20}
I.~Castillo and L.~B. Freidovich, ``Describing-function-based analysis to tune parameters of chattering reducing approximations of sliding mode controllers,'' {\em Control Engineering Practice}, vol.~95, p.~104230, 2020.

\bibitem{Lee07}
H.~Lee and V.~Utkin, ``Chattering suppression methods in sliding mode control systems,'' {\em Annual reviews in control}, vol.~31, no.~2, pp.~179--188, 2007.

\bibitem{Rosales15}
A.~Rosales and I.~Boiko, ``Disturbance attenuation for systems with second-order sliding modes via linear compensators,'' {\em IET Control Theory \& Applications}, vol.~9, no.~4, pp.~526--537, 2015.

\bibitem{Fridman01}
L.~Fridman, ``An averaging approach to chattering,'' {\em IEEE Transactions on Automatic Control}, vol.~46, no.~8, pp.~1260--1265, 2001.

\bibitem{Boiko07b}
I.~Boiko, L.~Fridman, A.~Pisano, and E.~Usai, ``Analysis of chattering in systems with second-order sliding modes,'' {\em IEEE transactions on Automatic control}, vol.~52, no.~11, pp.~2085--2102, 2007.

\bibitem{Boiko05}
I.~Boiko, ``Oscillations and transfer properties of relay servo systems—the locus of a perturbed relay system approach,'' {\em Automatica}, vol.~41, no.~4, pp.~677--683, 2005.

\bibitem{Boiko09}
I.~Boiko, {\em Discontinuous control systems: frequency-domain analysis and design}.
\newblock Boston: Birkh\"{a}user, 2009.

\bibitem{Gelb68}
A.~Gelb and W.~E. Vander~Velde, {\em Multiple-input describing functions and nonlinear system design}.
\newblock New York: McGraw-Hill, 1968.

\bibitem{Boiko04}
I.~Boiko, L.~Fridman, and M.~Castellanos, ``Analysis of second-order sliding-mode algorithms in the frequency domain,'' {\em IEEE Transactions on Automatic Control}, vol.~49, no.~6, pp.~946--950, 2004.

\bibitem{ByF05}
I.~Boiko and L.~Fridman, ``Analysis of chattering in continuous sliding mode controllers,'' {\em IEEE Transactions on Automatic Control}, vol.~50, no.~9, pp.~1442--1446, 2005.

\bibitem{Rosales17}
A.~Rosales, Y.~Shtessel, L.~Fridman, and C.~Panathula, ``Chattering analysis of hosm controlled systems: frequency domain approach,'' {\em IEEE Transactions on Automatic Control}, vol.~62, no.~8, pp.~4109--4115, 2017.

\bibitem{Ventura18}
U.~P{\'e}rez-Ventura and L.~Fridman, ``When is it reasonable to implement the discontinuous sliding-mode controllers instead of the continuous ones? frequency domain criteria,'' {\em International Journal of Robust and Nonlinear Control}, vol.~29, no.~3, pp.~810--828, 2018.

\bibitem{Ventura19}
U.~P{\'e}rez-Ventura and L.~Fridman, ``Design of super-twisting control gains: A describing function based methodology,'' {\em Automatica}, vol.~99, pp.~175--180, 2019.

\bibitem{Ventura21}
U.~P{\'e}rez-Ventura, J.~Mendoza-Avila, and L.~Fridman, ``Design of a proportional integral derivative-like continuous sliding mode controller,'' {\em International Journal of Robust and Nonlinear Control}, 2021.

\bibitem{Utkin20}
V.~Utkin, A.~Poznyak, Y.~Orlov, and A.~Polyakov, ``Conventional and high order sliding mode control,'' {\em Journal of the Franklin Institute}, vol.~357, no.~15, pp.~10244--10261, 2020.

\bibitem{LyF10}
A.~Levant and L.~Fridman, ``Accuracy of homogeneous sliding modes in the presence of fast actuators,'' {\em IEEE Transactions on Automatic Control}, vol.~55, pp.~810--814, March 2010.

\bibitem{Boiko08}
I.~Boiko, M.~Castellanos, and L.~Fridman, ``Analysis of response of second-order sliding mode controllers to external inputs in frequency domain,'' {\em International Journal of Robust and Nonlinear Control: IFAC-Affiliated Journal}, vol.~18, no.~4-5, pp.~502--514, 2008.

\bibitem{Boiko07}
I.~Boiko, M.~Castellanos, and L.~Fridman, ``Describing function analysis of second-order sliding mode observers,'' {\em International Journal of Systems Science}, vol.~38, no.~10, pp.~817--824, 2007.

\bibitem{Khalil92}
H.~K. Khalil, {\em Nonlinear Systems}.
\newblock New York: Macmillan, 1992.

\bibitem{Apkarian00}
P.~Apkarian and R.~J. Adams, ``Advanced gain-scheduling techniques for uncertain systems,'' in {\em Advances in linear matrix inequality methods in control}, pp.~209--228, SIAM, 2000.

\bibitem{Atherton75}
D.~P. Atherton, {\em Nonlinear control engineering}.
\newblock London: Van Nostrand Reinhold, 1975.

\bibitem{Booton54}
R.~C. Booton, ``Nonlinear control systems with random inputs,'' {\em IRE Transactions on Circuit Theory}, vol.~1, no.~1, pp.~9--18, 1954.

\bibitem{Loeb54}
J.~Loeb, ``Recent advances in nonlinear servo theory,'' {\em Transactions of the American Society of Mechanical Engineers}, vol.~76, no.~8, pp.~1281--1288, 1954.

\bibitem{Boiko18b}
I.~M. Boiko, ``On loeb's criterion of orbital stability of self-excited periodic motions,'' in {\em 2018 15th International Workshop on Variable Structure Systems (VSS)}, pp.~464--469, IEEE, 2018.

\bibitem{Martinez2021}
C.~A. Martínez-Fuentes, U.~Pérez-Ventura, and L.~Fridman, ``Chattering analysis for lipschitz continuous sliding-mode controllers,'' {\em International Journal of Robust and Nonlinear Control}, vol.~31, no.~9, pp.~3779--3794, 2021.

\end{thebibliography}


\end{document}